%% file: main.tex
\pdfoutput=1
\documentclass[reprint,amsmath,amssymb,aps]{revtex4-1}
\usepackage{graphicx}
\usepackage{dcolumn}
\usepackage{bm}
\usepackage{algorithm}
\usepackage{algpseudocode}
\usepackage{float}

\usepackage[caption=false,listofformat=subsimple,labelformat=simple]{subfig}

\usepackage[justification=raggedright]{caption}
\usepackage[export]{adjustbox}

\usepackage{textcomp} 

\usepackage{tabu}

\usepackage{color}
\definecolor{darkred}{RGB}{150,0,0}

\usepackage{xparse}
\RenewDocumentCommand\log{g}{%
  \IfNoValueF{#1}{\text{log}\left(#1\right)}%
  \IfValueF{#1}{\text{log}}%
}
\usepackage{hyperref}
\newcommand{\etal}{\textit{et al.}}

\begin{document}


\title{Convolutional restricted Boltzmann machine aided Monte Carlo: An application to Ising and Kitaev models}%

\author{Daniel Alcalde Puente}
\author{Ilya M. Eremin}%
\affiliation{%
 Institut f\"ur Theoretische Physik III,
		Ruhr-Universit\"{a}t Bochum, D-44780 Bochum, Germany
}%


\date{October 20, 2020}

\begin{abstract}
\input{abstract.tex}
\end{abstract}

\maketitle

\section{Introduction}
\input{introduction.tex}

\section{From fully connected RBM to Convolutional RBM}
\input{RBM.tex}

\section{A primer: 2D Ising Model}
\input{ising.tex}

\section{Kitaev Model on Honeycomb lattice}
\input{kitaev_math.tex}

\input{complexity.tex}

\input{kitaev.tex}
\section{Conclusion}
\input{conclusion.tex}
\bibliography{literature.bib}
\newpage
\begin{widetext}
\appendix
\section{Training}
\input{Atraining.tex}

\section{Sampling}
\input{Asampling.tex}

\section{Analytical solution}
\input{Aanalytical.tex}

\section{Training and Sampling Parameters}
\input{Atrainingandsampling}

\end{widetext}
\end{document}

%% file: abstract.tex
Machine learning is becoming widely used in analyzing the thermodynamics of many-body condensed matter systems. Restricted Boltzmann machine (RBM) aided Monte Carlo simulations have sparked interest recently, as they manage to speed up classical Monte Carlo simulations. Here we employ the Convolutional restricted Boltzmann machine (CRBM) method and show that its use helps to reduce the number of parameters to be learned drastically by taking advantage of translation invariance. Furthermore, we show that it is possible to train the CRBM at smaller lattice sizes, and apply it to larger lattice sizes. To demonstrate the efficiency of CRBM we apply it to the paradigmatic Ising and Kitaev models in two-dimensions.

%% file: introduction.tex
In recent years, machine learning and specifically restricted Boltzmann machines (RBMs) have been used in the field of many-body quantum systems to explicitly parametrize a probability distribution function of a quantum many-body state~\cite{Carrasquilla2020}. For example, RBMs have been found to be a powerful tool to obtain the ground state of quantum models through variational Monte Carlo (MC) \cite{carleo2017solving} and to reconstruct quantum states from a set of qubit measurements \cite{torlai2018neural, PhysRevB.100.195125}. RBMs have also been used to construct the exact ground state of quantum systems by reproducing the exact imaginary time evolution \cite{carleo2018constructing}. Neural networks have been shown to be useful in the classification of phases of matter in MC simulations \cite{carrasquilla2017machine}, which has awakened interest in the classification of topological and nematic phases \cite{Zhang2017QSL, PhysRevB.99.060404, rodriguez2019identifying}, as finding a suitable order parameter for phase transitions can be challenging. Increasing the resolution of already sampled states \cite{efthymiou2019super} has been another area of study. 
Another promising avenue in the application of the machine learning techniques is sizable speeding up of the quantum MC simulations obtained via the so-called self-learning Monte Carlo (SLMC) method \cite{liu2017self, xu2017self, shen2017self,Nagai2017self,Chen2018symmetry} applied to the models of interacting fermionic systems where the Trotter decomposition can be employed. In this method, the effective energy is inexpensive to compute and is mostly composed of two-particle interactions, which enables cluster updates. The strength of these two-particle interactions is then learned by applying the linear regression method. For more complex systems neural networks are employed to model the effective energy \cite{shen2018self,li2019accelerating}. This, however, makes cluster updates very hard to realize. At the same time, RBMs can be used as an alternative to SLMC for models where no Trotter decomposition is needed because the former can model more complex interactions than the original SLMC (learned by linear regression), and they are faster to sample from than SLMC with neural networks due to Gibbs sampling.

The RBM is a probabilistic model that has two main features. First, it is possible to sample states from the model's probability distribution with global updates (Gibbs sampling). In addition, the non-normalized probability distribution $\hat{P}_\text{RBM}(x;W)$ is well defined. Here, the matrix $W$ determines the actual form of the probability distribution and the task is to choose a $W$ such that $P_\text{RBM}(x;W)\approx P_\text{target}(x)$ for all $x$ and then use its global updates for sampling. This has been done for example for the Ising and Falikov-Kimball model \cite{torlai2016learning,huang2017accelerated} for small lattice sizes, $L=8$ and $L=10$. However, the training appears to be slow and major difficulties arise for RBM training for larger lattices.

One of the drawbacks of using RBMs in its current form is that for their training a Metropolis aided Monte Carlo simulation, which is expensive, needs to be performed in advance. A well-behaving Metropolis simulation is needed to train an RBM, which poses a problem since the scenario where Metropolis is not behaving well is exactly the one where we want to apply the RBM. The second drawback is that for larger lattice size $L$, the number of parameters that need to be learned scale with $L^4$ in two-dimensions. So the bigger $L$ is, the more training time is required.

To overcome these problems, we propose to use Convolutional restricted Boltzmann machines (CRBMs), which utilizes translation invariance and was originally developed in the context of feature extraction in images \cite{norouzi2009stacks,lee2009convolutional}. In this paper, we apply CRBMs to the paradigmatic Ising and Kitaev models in two-dimensions. Using the fact that CRBMs are translationally invariant we demonstrate that using them reduces the number of parameters that need to be learned, which leads to faster training. In addition, the same CRBM can be applied to different lattice sizes, which means that after the CRBM has learned a probability distribution with a certain lattice size $L_\text{small}$, it can then be scaled to larger lattice sizes without extra computational costs. Furthermore, we show that if the CRBMs does not fit the probability distribution exactly, it can be corrected using a version of parallel tempering. Convolutions have been employed before in the context of Neural Network SLMC \cite{shen2018self,li2019accelerating} but after the convolution a fully connected layer is applied, which breaks translations invariance. Note that CRBM can only be applied to statistical mechanics models where translation invariance is preserved. This, for example, rules out its application to random spin systems.

The paper is organized as follows. In the next section, we present the details of the CRBM and in Sec.~\ref{sec:results_ising} the autocorrelation times are compared between the Metropolis MC and CRBM for the Ising model. We extend the results to the two-dimensional 2D Kitaev model in Sec.~\ref{sec:results_kitaev} where the specific heat at different temperatures is computed for both periodic and open boundary conditions. In this section, the CRBM is also compared to the fully connected RBM. The main results are summarized in the Conclusion.

%% file: RBM.tex
\begin{figure}[!ht]
\centering 
\subfloat[]{\label{fig:weightsfc}\includegraphics[width=0.5\textwidth]{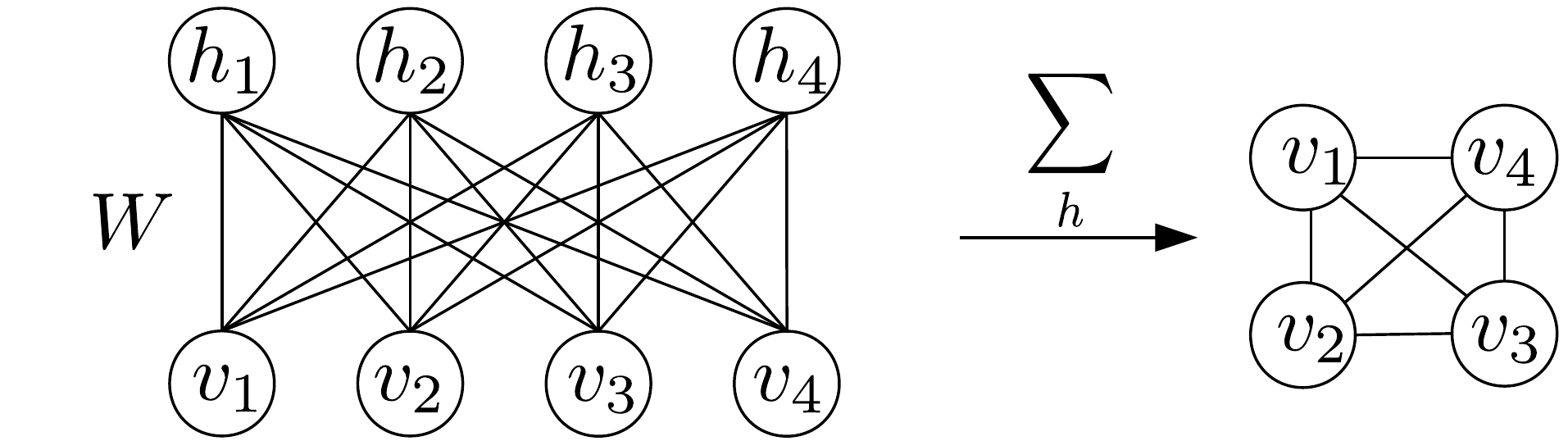}}\\
\subfloat[]{\label{fig:weightsc}\includegraphics[width=0.5\textwidth]{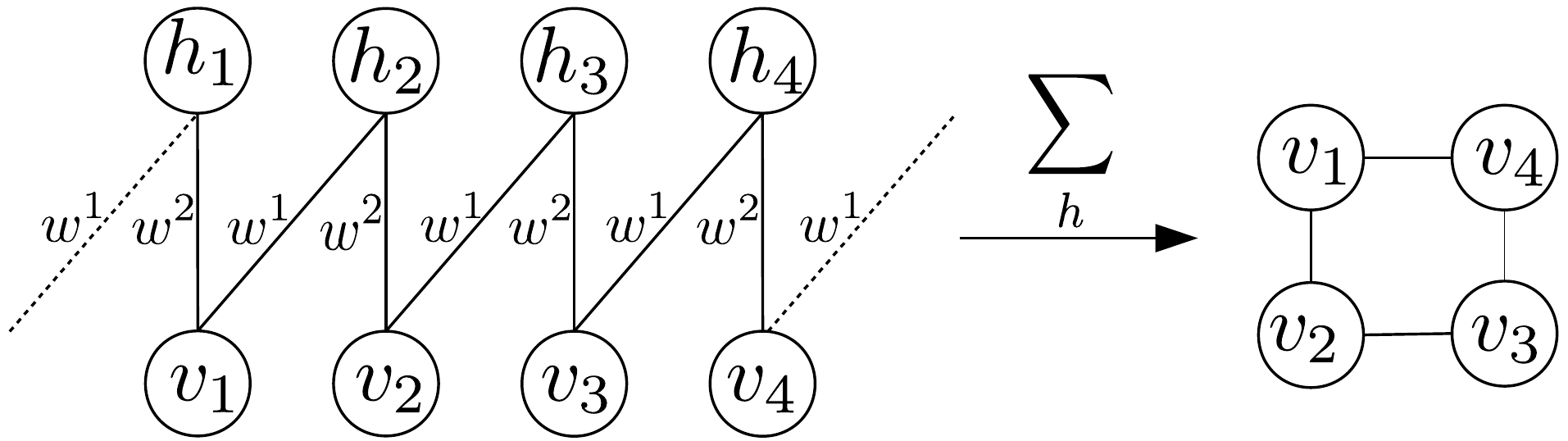}}
\caption{One-dimensional representation of (a) fully connected RBMs and (b) CRBMs. Left panels show a connection between visible and hidden units. On the right panel the effective connections between visible units after summing over $h$, representing the free energy (see Eq.~\ref{eq:rbm_free-energy}) are shown. Observe that only nearest neighbors are connected through $h$ in the CRBM and so only nearest neighbors interact.}
\end{figure}

A restricted Boltzmann machine (RBM) is a probabilistic generative neural network model, which can be used to learn an approximate probability distribution and then sample from it using the Gibbs sampling. 
The model has two distinct groups of statistical variables, the visible variables $v$ and the hidden variables $h$ as shown in Fig.~\ref{fig:weightsfc}. Here, $v$ and $h$ are the vectors of length $N$ where each component takes either the value 0 or 1, i.e. $v \in \{0,1\}^{N_p}$ and $h \in \{0,1\}^{N_h}$, where $N_p$ is the size of the vector $v$ and $N_h$ is the size of the vector $h$. Note that for a square lattice $N_p=L^2$. The probability distribution over both the visible and the hidden variables is $P_\text{RBM}(v,h;W)$. Summing over the hidden variables, the probability distribution over the visible units is given as $P_\text{RBM}(v;W)=\sum_h P_\text{RBM}(v,h;W)$, which after training will approximate the target distribution. The hidden units are required as instruments for Gibbs sampling and to mediate the interaction between visible units. Visible and hidden variables form a bipartite system, connected through a matrix $W$ to each other but not to themselves. The probability distribution over both visible and hidden variables can be expressed as:
\begin{align}
\label{eq:rbm_energy}
P_\text{RBM}(v,h)&=\frac{e^{-E(v,h)}}{Z} \\
E(v,h)&=-\sum_i h_\text{bias}^i h^i -\sum_j v_\text{bias}^j v^j - \sum_{ij} h^i W_{ij} v^j
\end{align}
where $Z$ is the normalization constant and $W$, $v_\text{bias}$, and $h_\text{bias}$ are the model parameters. 
Summing over $h$, the visible units $v_i$ are no longer independent of each other (see Fig.~\ref{fig:weightsfc}) in the sense that each $h_j$ now represents an interaction of the $v_i$ that are connected via $W_{ij}$. The probability over the visible units is then given by:
\begin{align}
\label{eq:rbm_free-energy}
P_\text{RBM}(v)&=\sum_{h \in \{0,1\}^N} P(v,h) = \frac{e^{-F(v)}}{Z} \\
F(v)&=-\sum_j v_\text{bias}^j v^j - \sum_{i} \log{1 + e^{h_\text{bias}^i +\sum_j W_{ij} v^j}}.
\end{align}
The physical model with energy $E_\text{phys}(x)\beta$, will then be approximated by the energy $F_\text{RBM}(x;W)$ of the RBM.
Training is then done supervised similarly to Ref.~\cite{huang2017accelerated} by minimizing the loss:
\begin{align}
\label{eq:loss}
\text{loss}(W) = \frac{1}{M}\sum_{i=1}^M [E_\text{phys}(x_i)\beta - F_\text{RBM}(x_i;W) - C(W)]^2
\end{align}
where $\beta$ is the inverse temperature and $C$ is a value that can be chosen freely since the probability function is invariant under the addition of a constant to the energy. Note that if $\text{loss}(W)=0$ than $F_\text{RBM}(x_i;W)=E_\text{phys}(x_i)\beta + C(W)$ and so $P_\text{RBM}(x;W)=P_\text{phys}(x)$. $C$ is chosen such that the loss is minimal:
\begin{align}
C(W) = \frac{1}{M} \sum_{i=1}^M E_\text{phys}(x_i)\beta - F_\text{RBM}(x_i;W).
\end{align}
Observe that this minimization of the loss is equivalent to minimizing the Kullback-Leibler divergence:
\begin{align}
D_\text{KL}(P_\text{RBM}||P_\text{phys})=\sum_{x\in X}P_\text{RBM}(x;W)\log{\frac{P_\text{RBM}(x;W)}{P_\text{phys}(x)}}.
\end{align}
\begin{widetext}
In particular, It can be rewritten as:
\begin{align}
D_\text{KL}(P_\text{RBM}||P_\text{phys})=\sum_{x\in X}P_\text{RBM}(x;W)\left[ E_\text{phys}(x)\beta - F_\text{RBM}(x)\right]+\log{Z_\text{phys}}-\log{Z_\text{RBM}}, 
\end{align}
and its gradient is given by:
\begin{align}
\frac{\partial D_\text{KL}(P_\text{RBM}||P_\text{phys})}{\partial W}=
\biggl\langle
\left[\biggl \langle \frac{\partial F(x;W)}{\partial W}\biggr \rangle_{x\in P_\text{RBM}(x)} -  \frac{\partial F(x;W)}{\partial W} \right]\left[ E_\text{phys}(x)\beta - F_\text{RBM}(x;W)\right] 
\biggr\rangle_{x\in P_\text{RBM}(x)},
\end{align}
which is equivalent to performing the derivative on our loss:
\begin{align}
\biggl\langle
\frac{\partial
\frac{1}{2}(E_\text{phys}(x)\beta - F_\text{RBM}(x;W) - C(W))^2
}
{\partial W}
\biggr\rangle_{x\in P_\text{RBM}(x)}.
\end{align}
\end{widetext}
Note that in practice, we use a mixture of states sampled from both the physical probability distributions and the RBM's probability distributions.
Also, note that this version of the Kullback-Leibler divergence will converge faster than the one commonly used in machine learning for unsupervised learning $D_\text{KL}(P_\text{phys}||P_\text{RBM})$. In our application, we can use it since we have prior knowledge of the energy of the target distribution, which is not the case for unsupervised machine learning tasks.
Note that ADAM batch-gradient descent \cite{kingma2014adam} is used to minimize this loss.
The only remaining question is, which states $x_i$ do we train on?
The authors of Ref. \cite{huang2017accelerated} have proposed to train with the states sampled from the physical distribution.
The problem with this is that if a state $x’$ has a tiny realization probability $P_\text{phys}(x’)\ll 1$, then the RBM cannot learn the state $x’$ due to its rare realization. This leaves low probability regions of $P_\text{phys}$ undefined. This is especially a problem at low temperatures.
Therefore, the solution to the problem should be not only to train RBM with samples from $P_\text{phys}$ but also to include samples from the RBM itself. This will suppress the development of high probability regions of $P_\text{RBM}$ in areas where no physical states are sampled. In practice, two steps have to be performed.
First, states from the physical distribution are sampled through the Metropolis algorithm. Second, in each training step, we sample from the RBM using Gibbs sampling (see~App.~\ref{sec:gibbs}), combine it with samples obtained previously with Metropolis MC, and then perform one training step. In addition, a pre-training step is also performed. In this step, the CRBM is only trained using random samples as discussed in App.~\ref{sec:Atraining}.
\begin{figure}
	\centering
	\includegraphics[width=0.5\textwidth]{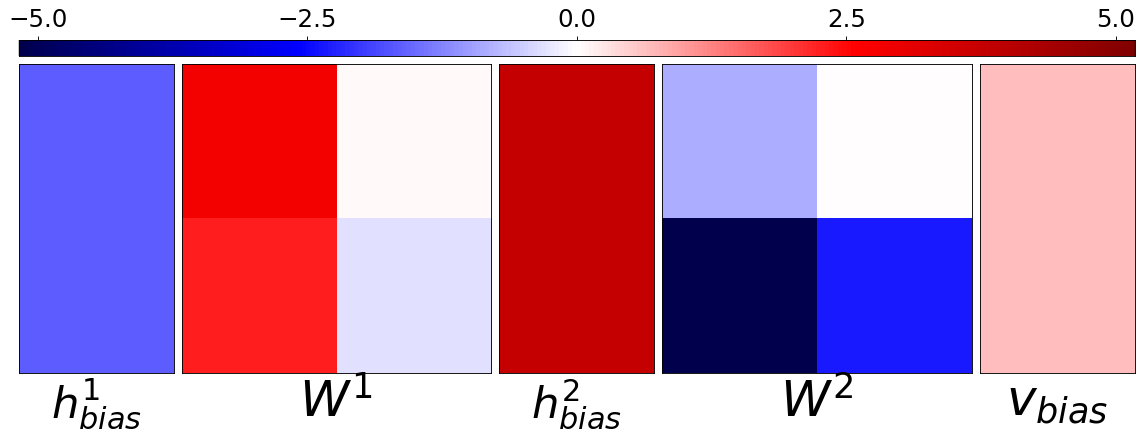}
	\caption{Convolution kernels $W_k$ (see Eq.~\ref{eq:free-energy_CRBM}) for the CRBM with two convolutional kernels, two hidden bias, and one visible bias. The CRBM was trained at $T=2.2$.}
	\label{fig:ising_CRBM_kernel}
\end{figure}
The biggest drawback of using the RBM in Monte Carlo simulation is that a Metropolis simulation (numerically expensive) needs to be performed before training. This usually makes the use of RBMs redundant since they can only be trained if the Metropolis algorithm performs well in which case expectation values could have been computed directly. The second drawback as mentioned in the Introduction is that for larger lattice size $L$, the number of parameters that need to be learned scales with $L^4$ in two-dimensions. So the larger $L$ is, the more training time is required.
To overcome these difficulties, we propose to employ Convolutional RBM (CRBM), a model used originally in image feature extraction \cite{norouzi2009stacks,lee2009convolutional}, which exploits the fact that models in question are translationally invariant and the interactions are local. In particular, the matrix $W$ can be chosen such that the probability distribution $P_\text{RBM}(v;W)$ is translationally invariant. Consequently, instead of all lattice points being connected to each other like in the fully connected RBM, in the CRBM only the neighboring lattice sites are connected.
\begin{figure*}
	\centering
	\subfloat[]{\label{fig:ising_CRBM_auto}\includegraphics[width=0.33\textwidth]{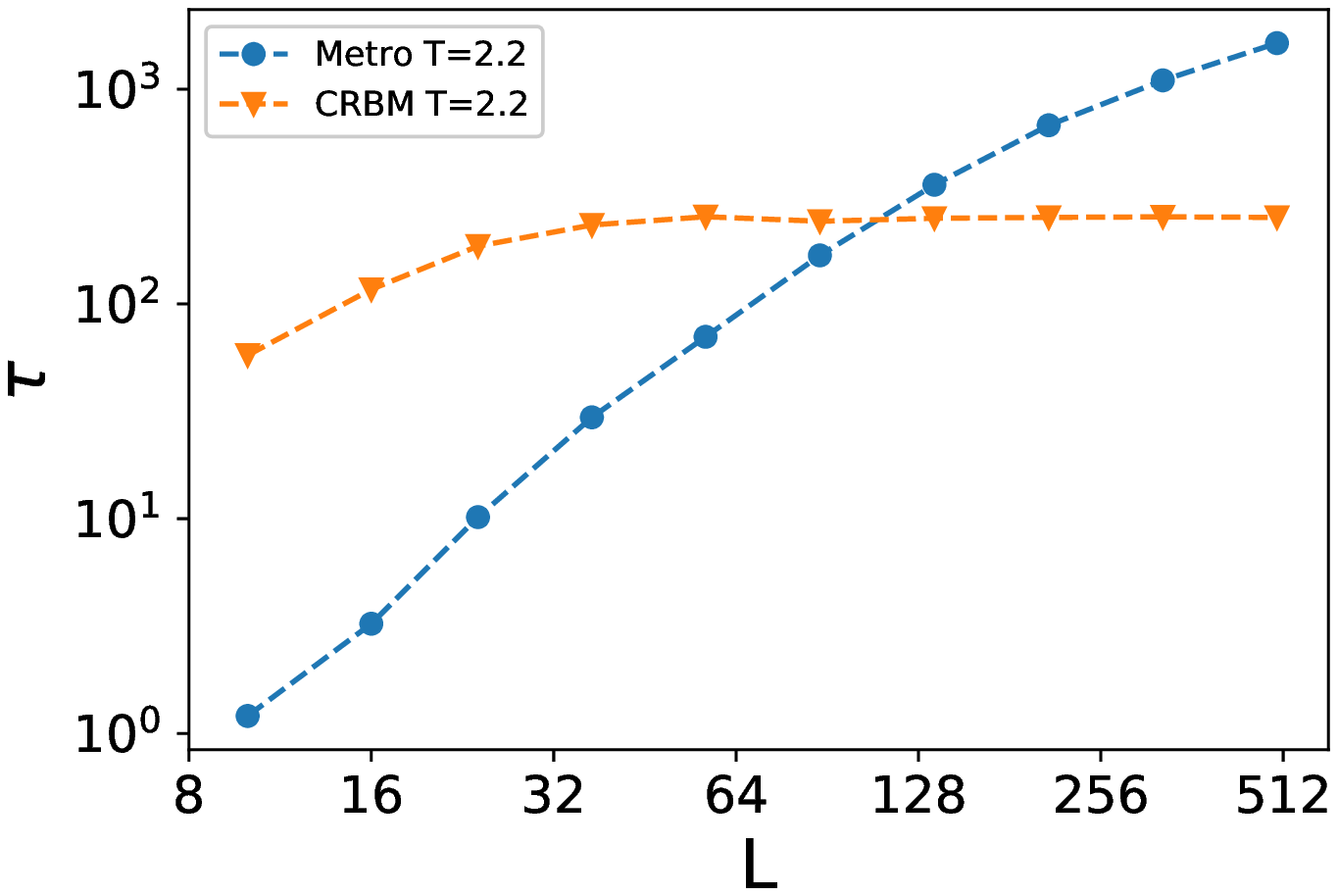}}
	\subfloat[]{\label{fig:ising_CRBM_auto2}\includegraphics[width=0.33\textwidth]{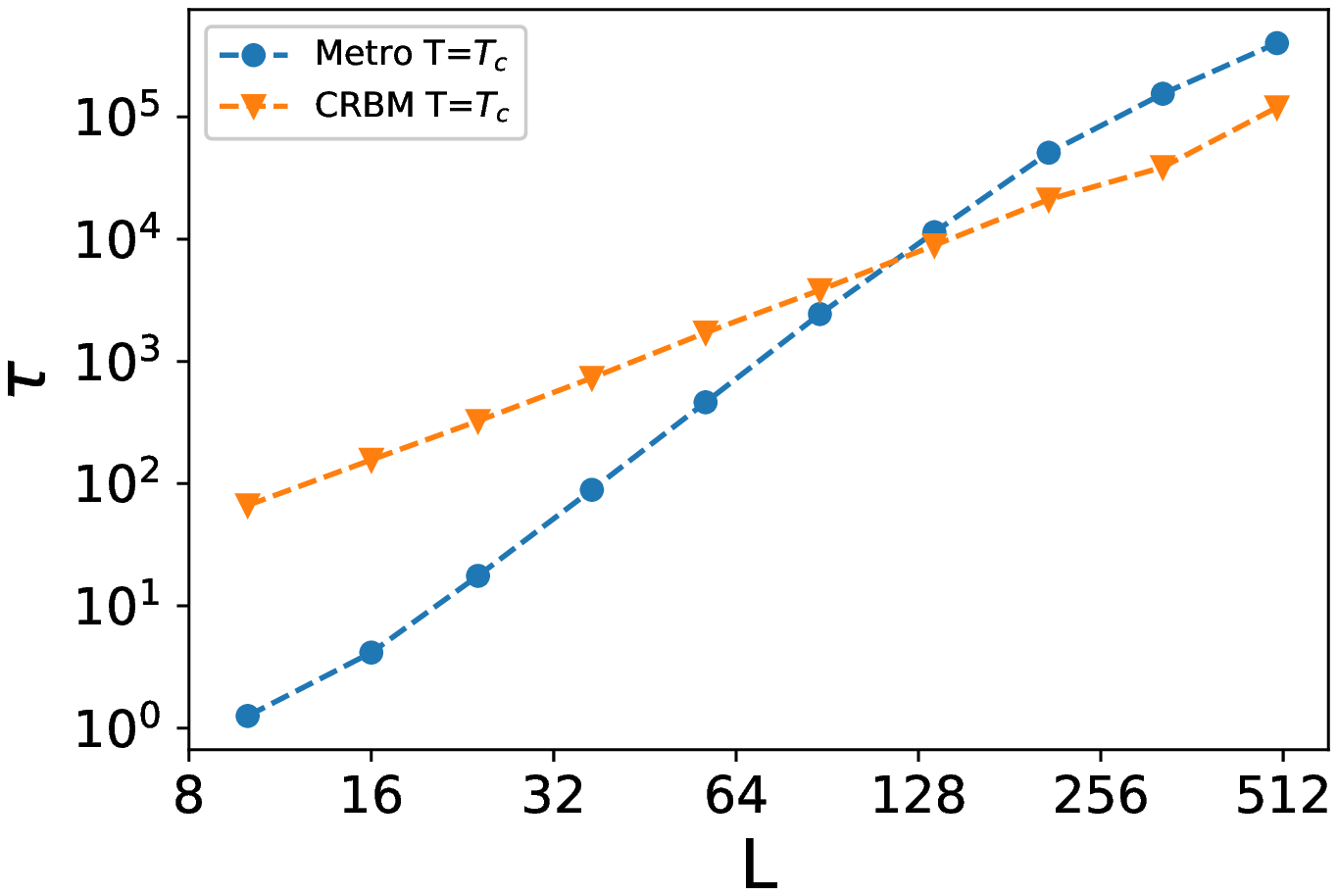}}\\
	\subfloat[]{\label{fig:ising_cv}\includegraphics[width=0.33\textwidth]{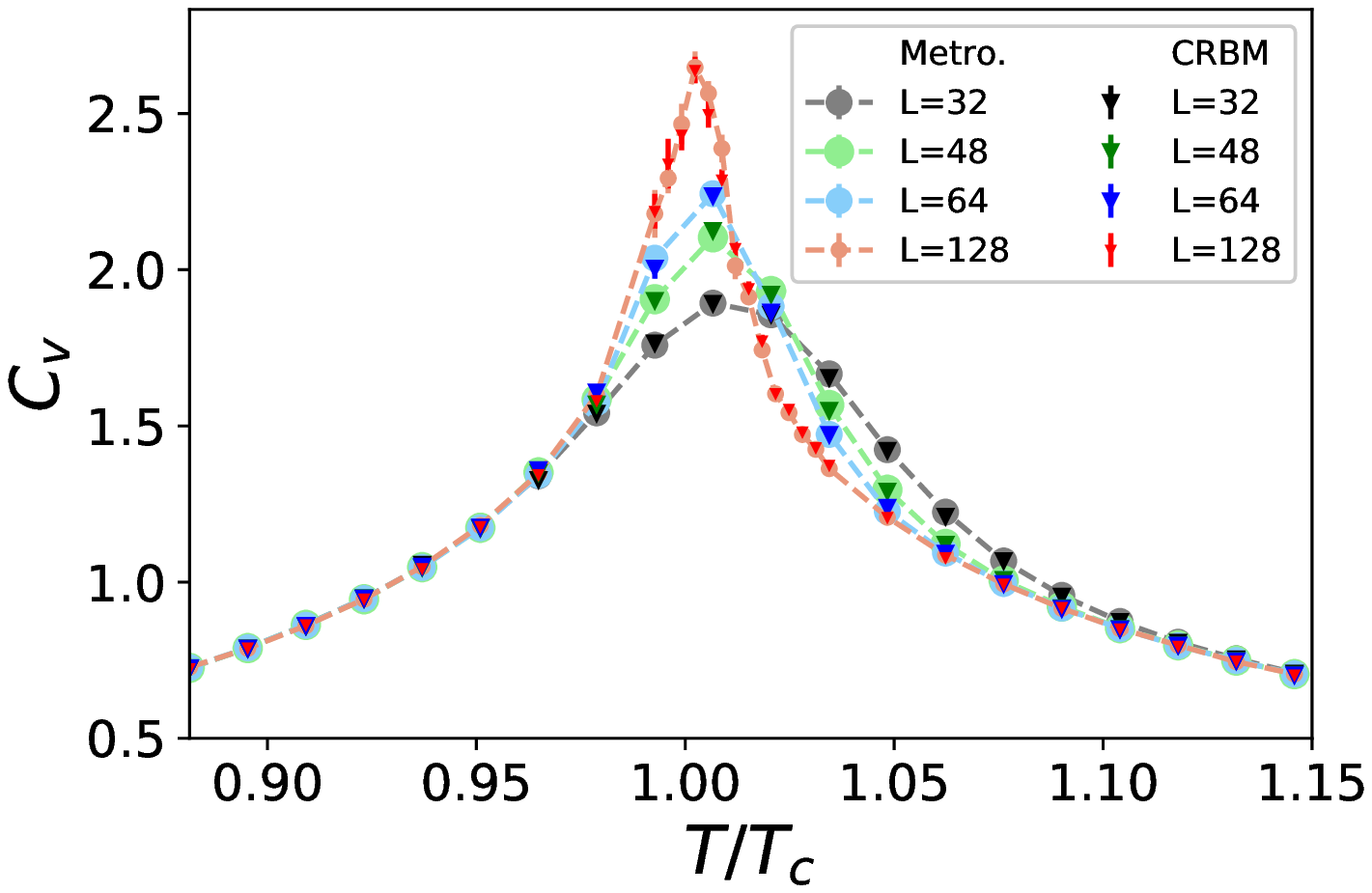}}
	\subfloat[]{\label{fig:ising_susceptibility}\includegraphics[width=0.33\textwidth]{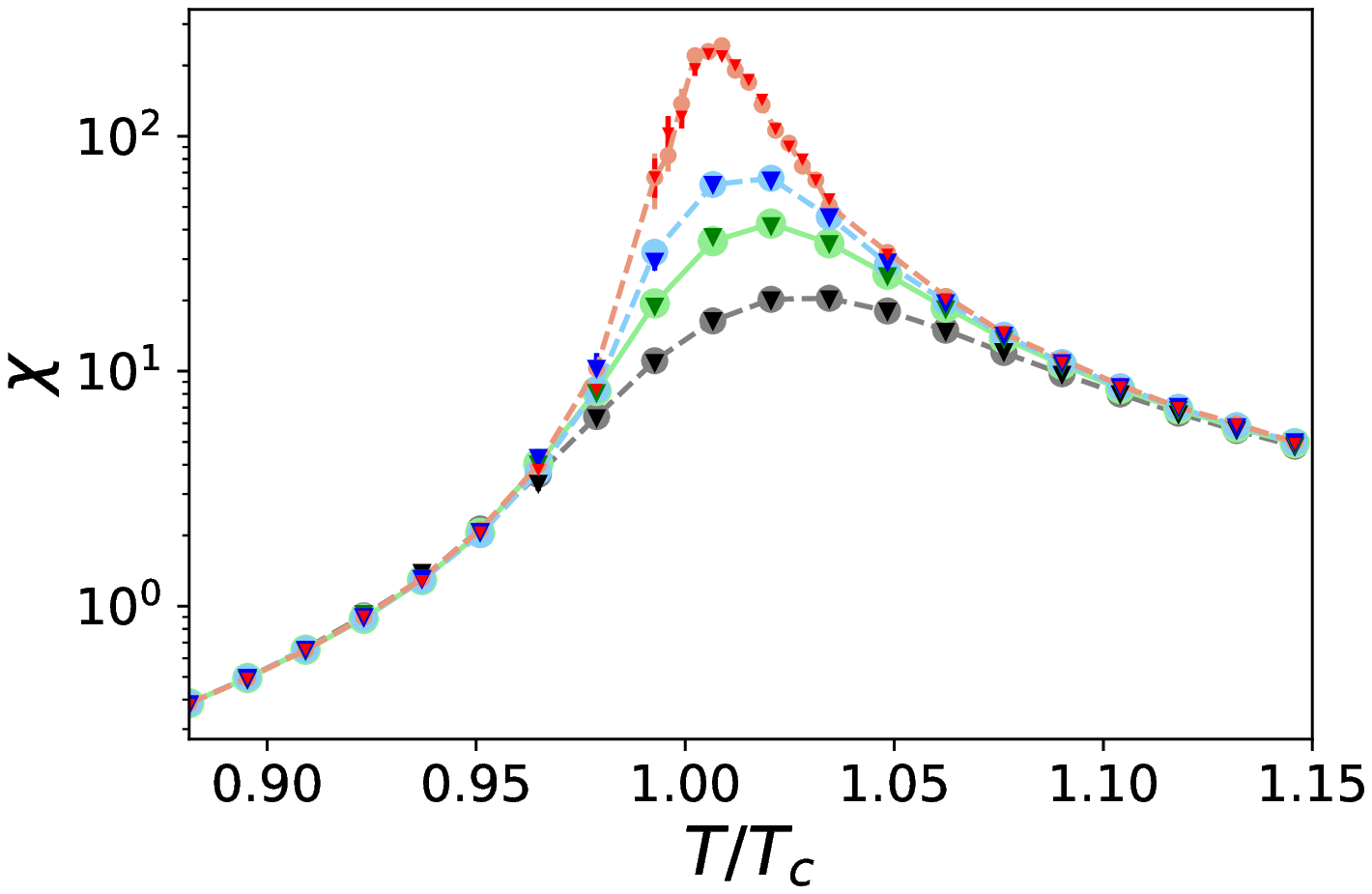}}
	\subfloat[]{\label{fig:ising_susceptibility_finite}\includegraphics[width=0.33\textwidth]{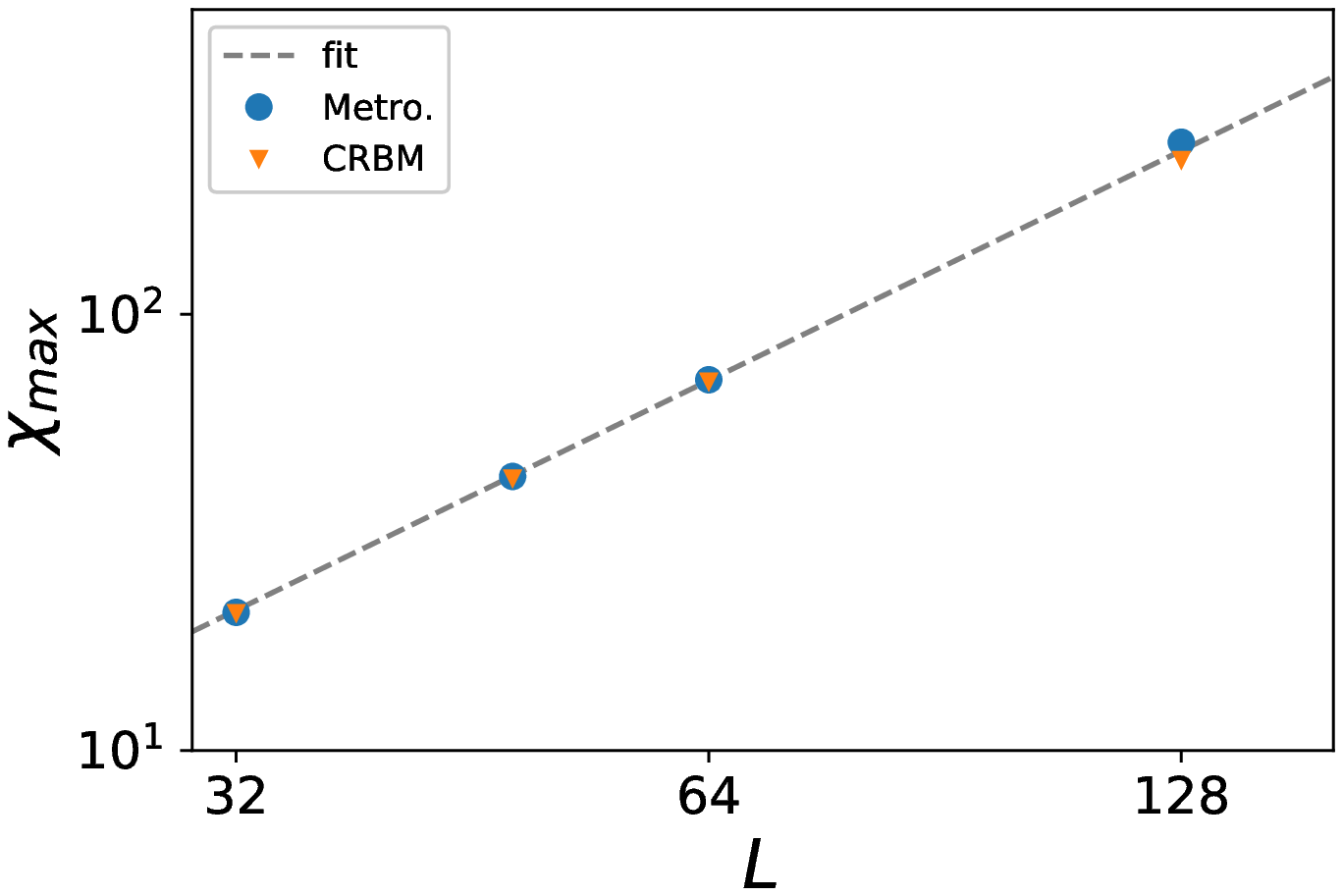}}
	\caption{
	Autocorrelation time at different lattice sizes at (a) $T=2.2$ and (b) $T=T_c$ for simulations with $10^8$ samples. (c) Specific heat and (d) magnetic susceptibility at $L=\{32,48,64,128\}$ for an MC with $10^7$ samples. (e) Finite size scaling of the maximum of the susceptibility at different lattice sizes yielding the expected critical exponent of $\gamma_\text{CRBM}=1.7$ and $\gamma_\text{metro.}=1.8$ vs the theoretical value of $\gamma_\text{theo.}=1.75$. For a higher accuracy in the critical exponent more samples at $L=128$ and a more rigorous finite size scaling would be necessary.
	}
\end{figure*}
This is illustrated in Fig.~\ref{fig:weightsc} for the 1D case. Translation symmetry also requires that $v_1$ should be connected to $v_2$ in the same way as $v_3$ is connected to $v_4$. With nearest-neighbor interactions, applying this condition, only two weights remain. After summing over $h$ an Ising like interaction is obtained.

The interaction part of the 'free' energy for Fig.~\ref{fig:weightsc} is:
\begin{align}
F_{\text{CRBM}}(v)&=f(v_1,v_2) + f(v_2,v_3) + f(v_3,v_4) + f(v_4,v_1) \nonumber \\
f(v_i,v_k;W)&=\log{1+e^{h+w^1 v_i + w^2 v_j}}.
\end{align}
If $W$ is chosen such that $f(v_i, v_j;W) = \beta J s_is_j+C$ then the RBM is equivalent to Ising model ($s_i=2v_i-1$). In App.~\ref{sec:Aanalytical}, an analytical mapping between RBM and the Ising model is calculated as in Ref.~\cite{IsingAnalytical}. This is a good point to emphasize that the RBM does not directly learn statistical correlations but learns how to reproduce the energy of the target distribution. Even though the statistics differ greatly between small latices and large lattice sizes, the structure of the energy remains unchanged and so the CRBM can be trained at a small lattice size and then applied to large lattice sizes.

For the two-dimensional case, the energy is:
\begin{align}
E(v,h) = &-\sum_{k,i,j} h^k_{ij} (W^k * v)_{ij} \label{eq:energy_CRBM}\\
&- \sum_{k} h_\text{bias}^k \sum_{i,j} h_{ij}^k - v_\text{bias} \sum_{i,j} v_{ij} \nonumber
\end{align}
where the symbol $*$ represents a wrap around convolution between the lattice and the kernel.
Note that the visible units have dimension $L\times L$ and the hidden units have dimension $K\times L\times L$. They are connected through $K$ convolutional kernels $W^k$.
After summing over the hidden units, this gives the free energy:
\begin{align}
\label{eq:free-energy_CRBM}
F(v) = -v_\text{bias} \sum_{i,j} v_{ij} -\sum_{k,i,j} \log{1+e^{(v*W^k)_{ij} +h^k_\text{bias}}}.
\end{align}

Note that the boundary conditions of the CRBM will be adapted to the ones of the physical problem.
In the following sections, we will compare Metropolis MC and CRBM results in the application to the Ising and the Kitaev model in two-dimensions.

%% file: ising.tex
\label{sec:results_ising}
The 2D Ising model with nearest-neighbor ferromagnetic interaction, $J<0$, among the spins $s_i$%
\begin{equation}
E(s)=-J\sum_{\langle ij\rangle}s_i s_j
\end{equation}
is one of the simplest models in numerical statistical physics and RBMs have been successfully applied \cite{torlai2016learning}. Here, the local update Metropolis is compared with the CRBM as a proof of concept. We employ Theano \cite{al2016theano}, a python library, for optimization and sampling. Differentiation of the loss is performed automatically by the library.
Calculations for the Metropolis-Hasting algorithm were performed using a CPU and the convolutions needed for the CRBM were computed using a GPU.

The CRBM consists of two kernels with size $2\times2$ and is first trained for the temperature $T=2.2$. The model has $2\times 2\times2$ weights, one visible bias, and two hidden biases. These are 11 free parameters for the free energy $F_\text{RBM}(x)$. The two trained kernels as expected only show interactions between nearest neighbors (Fig.~\ref{fig:ising_CRBM_kernel}). The CRBM can be trained with $L=3$ states since the interaction is only nearest neighbours. Furthermore, we stress that to train the CRBM for the Ising model no Metropolis MC simulation is necessary as the structure of the energy distribution is simple. At $L=3$ there are only $2^{3^2}=512$ possible states so instead of first sampling with Metropolis we can directly train the CRBM with those states or alternatively directly use the analytical mapping between the Ising model and the CRBM found in App.~C. At the same time, once the CRBM has been trained, the MC simulation to compute expectation values can be performed. The emergent statistical behavior of the simple structure of the energy leads to finite-size effects.

Samples are generated with both the Metropolis MC algorithm and the previously trained CRBM.
For the Metropolis algorithm, $k$ steps are performed between each recorded step, while only one Gibbs step is performed between each recorded step by the CRBM. $k$ is chosen such that both simulations take the same amount of time (see also App.~\ref{sec:A_parameters}).
For example at $L=500$, $k=10^4$ Metropolis steps take the same time as one CRBM step.
The behavior of the thermodynamic observables at different temperatures matches between Metropolis and CRBM (see~Fig.~\ref{fig:ising_cv} and \ref{fig:ising_susceptibility}). The autocorrelation of the energy is compared between different lattice sizes at $T=2.2$ (Fig.~\ref{fig:ising_CRBM_auto}) and $T=T_c=\frac{2}{\ln{(1+\sqrt{2})}}$. At small lattice sizes, the Metropolis MC algorithm performs better than the CRBM, which reverses at $100<L$. At $L=500$ and $T=2.2$ ($T=Tc$) the CRBM is 7 times (4 times) faster than Metropolis MC. It is important to realize that for the Ising model there are cluster update algorithms that would outperform the CRBM. The only remarkable thing is that for interactions where these techniques are not available, the CRBM can be used. The use of the CRBM will be shown to be especially useful in the application in systems where it is computationally expensive to compute the energy as will be shown in the next section. A Jupyter notebook with a showcase for the Ising model can be found in \cite{github}.

%% file: kitaev_math.tex
\begin{figure}
	\centering
	\includegraphics[width=0.25\textwidth]{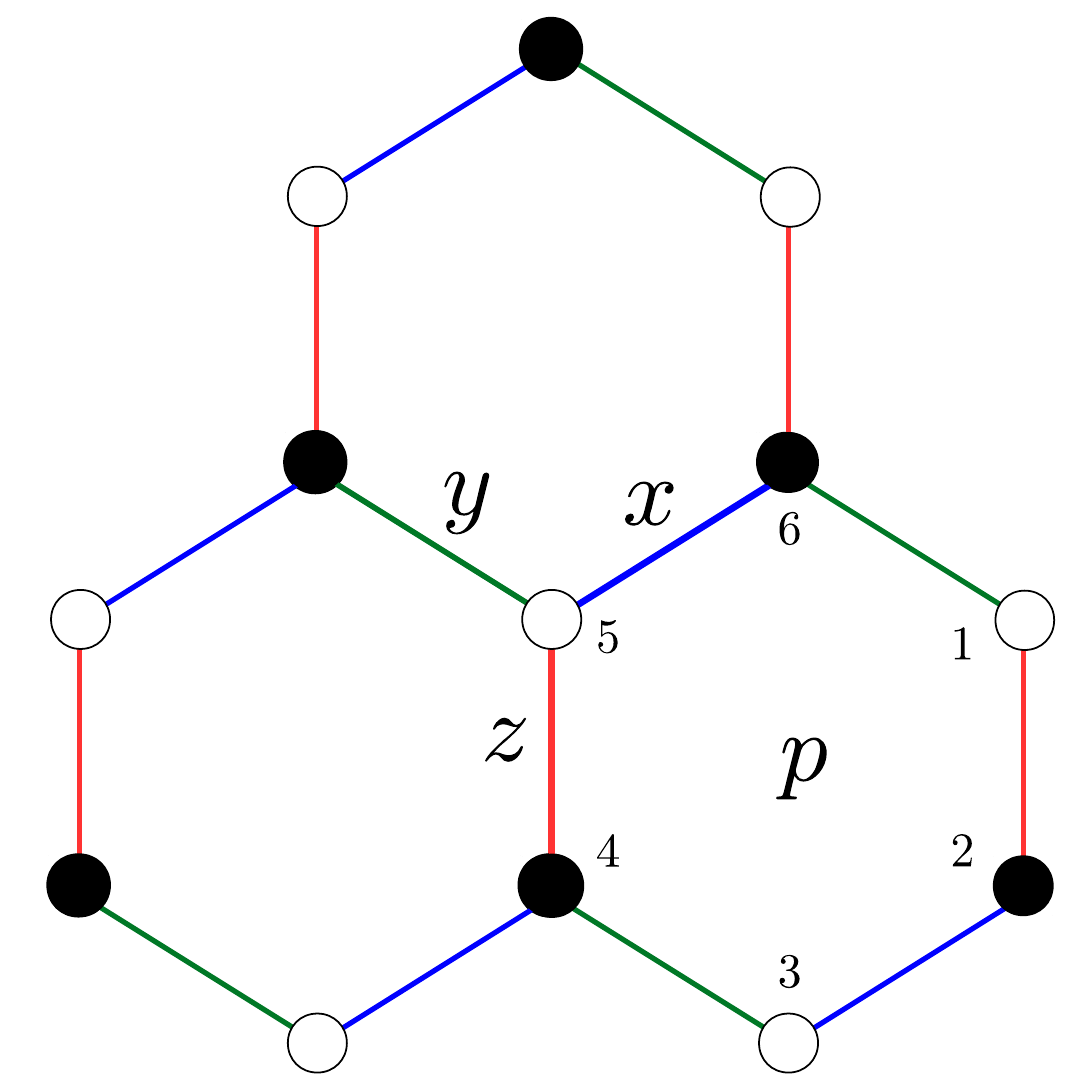}
	\caption{Illustration of the honeycomb lattice. The unit cell consists of two atoms (black and white dots).}
	\label{fig:honeycomb}
\end{figure}

\label{sec:results_kitaev}
Kitaev’s honeycomb lattice model \cite{kitaev} is the actual model to which we would like to apply the CRBM. It acquired significant attention recently due to the non-trivial spin liquid ground state with Majorana excitations \cite{2Dkitaev}, which can be computed analytically. The Hamiltonian on the honeycomb lattice as shown in Fig.~\ref{fig:honeycomb}, has the following form:
\begin{align}
H=-J_x\sum_{\langle ij \rangle_x} \sigma^x_i\sigma^x_j-J_y\sum_{\langle ij \rangle_y} \sigma^y_i\sigma^y_j-J_z\sum_{\langle ij \rangle_z} \sigma^z_i\sigma^z_j
\end{align}
with anisotropic exchange interaction, $J_x$, $J_y$, $J_z$ and Pauli matrices $\sigma^\alpha_{i}$.
Applying the Jordan-Wigner and the Majorana transformation \cite{2Dkitaev} the Hamiltonian transforms to:
\begin{align}
H=-iJ_x\sum_{x \text{ bonds}} c_b c_w - i J_y \sum_{y \text{ bonds}} c_b c_w - i J_z \sum_{z \text{ bonds}} \eta_r c_b c_w 
\end{align}
where $c_w$ ($c_b$) are Majorana operators defined at white (black) lattice points and $\eta_r=i\bar{c}_b \bar{c}_w=\pm 1$ are classical $\mathbb{Z}_2$ variables defined on each $z$ bond. Through eigenvalue decomposition of the Hamiltonian, one obtains a free energy $F(\eta_r)=-T \sum_\lambda \log{2\cosh(\beta \epsilon_\lambda/2)}$. The $\eta$ configuration will be sampled from $P(\eta_r)=\frac{e^{-\beta F(\eta_r)}}{Z}$. The temperature is in units of the interaction strengths $J_\alpha$ and $\hbar=k_\textup{B}=1$. The usual Metropolis MC simulation is performed following Ref.~\cite{nasu2014vaporization} using periodic boundary conditions in the z-direction. 
As proof of concept, we first employ CRBM using periodic boundary conditions for both z- and for x-y-directions. Next, we will explore the open boundary condition only along $z$-direction as in Ref.~\cite{nasu2014vaporization}. Note that for all our calculations the interaction strength is chosen to be $J_\alpha=\frac{1}{3}$.

It is important to note that the free energy of the Kitaev model is expensive to compute, which is in contrast to the Ising model were computing the energy is cheaper than one CRBM Gibbs step. For the Kitaev model, the computation of the free energy is as expensive as $k=4$ Gibbs updates at lattice size $L=8$ and as expensive as $k=350$ Gibbs updates at $L=30$. For more details see App.~\ref{sec:A_parameters}.

%% file: complexity.tex
Monte Carlo simulations of the Kitaev model have two major difficulties upon increasing the lattice size $L$. First, the computational complexity of the free energy scales with $\mathcal{O}(L^6)$. Second, the autocorrelation time $\tau_\text{metro}(L)$ of the Metropolis algorithm increases with larger $L$. In total, the complexity increase is $\mathcal{O}(\tau_\text{metro}(L)L^{6})$.

The CRBM tackles the slowness of the Metropolis algorithm. Instead of having to do expensive Metropolis steps, we employ cheap CRBM steps (convolution) $\mathcal{O}(L^{2})$ until the original state is uncorrelated to the new state. This new state is then corrected through the parallel tempering exchange correction. The correction step involves computing the expensive free energy $\mathcal{O}(L^{6})$. This means that if the CRBM is close enough to the physical distribution, this approach has a complexity $\mathcal{O}(L^{6} + \tau_\text{CRBM}(L) L^{2})$. 
So we conclude that a well trained CRBM will perform Monte Carlo with a complexity of $\mathcal{O}(L^{6})$ instead of $\mathcal{O}(\tau_\text{metro}(L)L^{6})$.
In the following, we apply CRBM to the Kitaev model with open and periodic boundary conditions.

%% file: kitaev.tex
\subsection{Periodic Boundary Conditions}
\subsubsection{Comparing FRBM and CRBM with Metropolis}
\label{sec:kitav_periodic}

\begin{figure}
    \centering
    \subfloat[]{\label{fig:kitaev_CRBM_kernel}\includegraphics[width=0.5\textwidth]{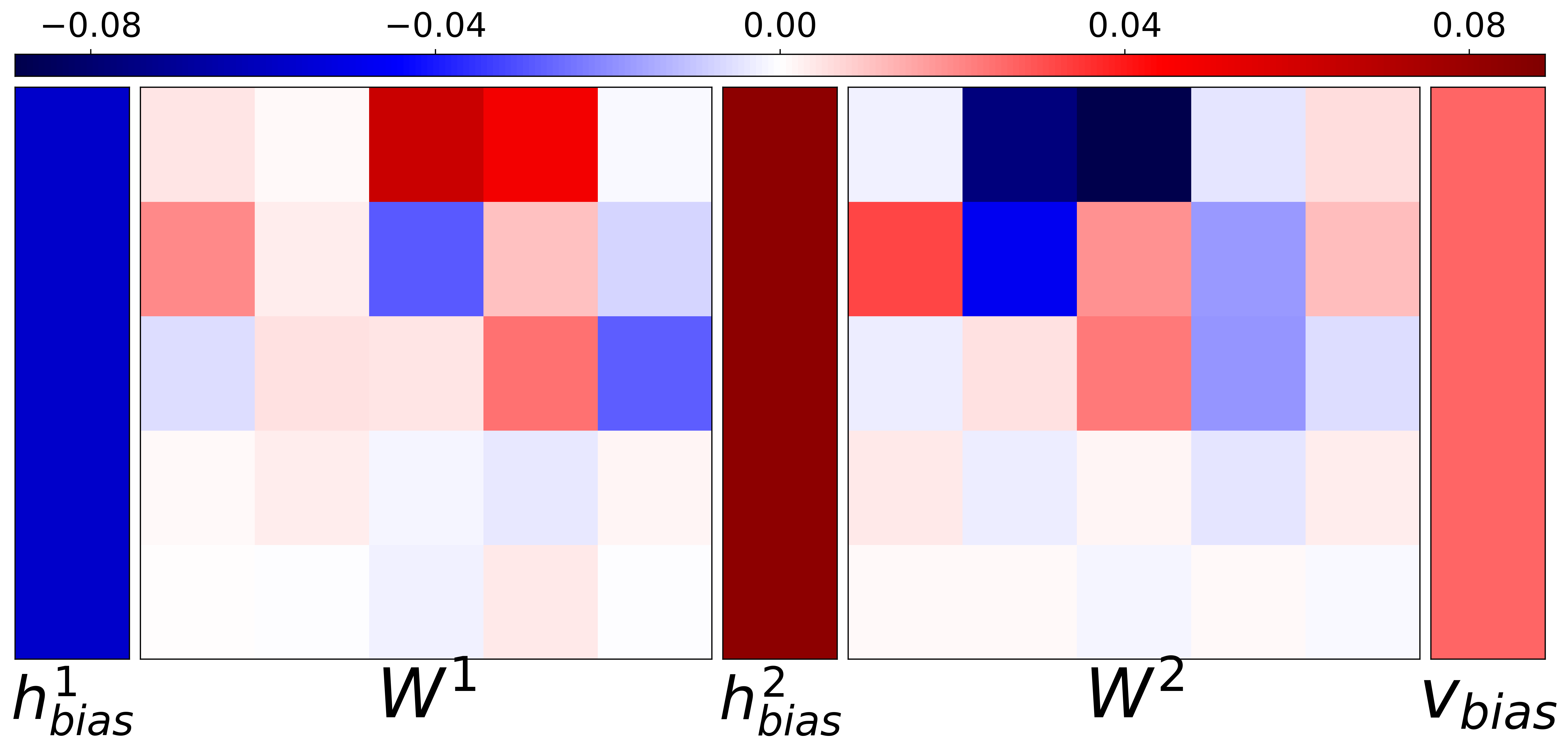}}\\
    \subfloat[]{\label{fig:kitaev_FRBM_kernel}\includegraphics[width=0.5\textwidth]{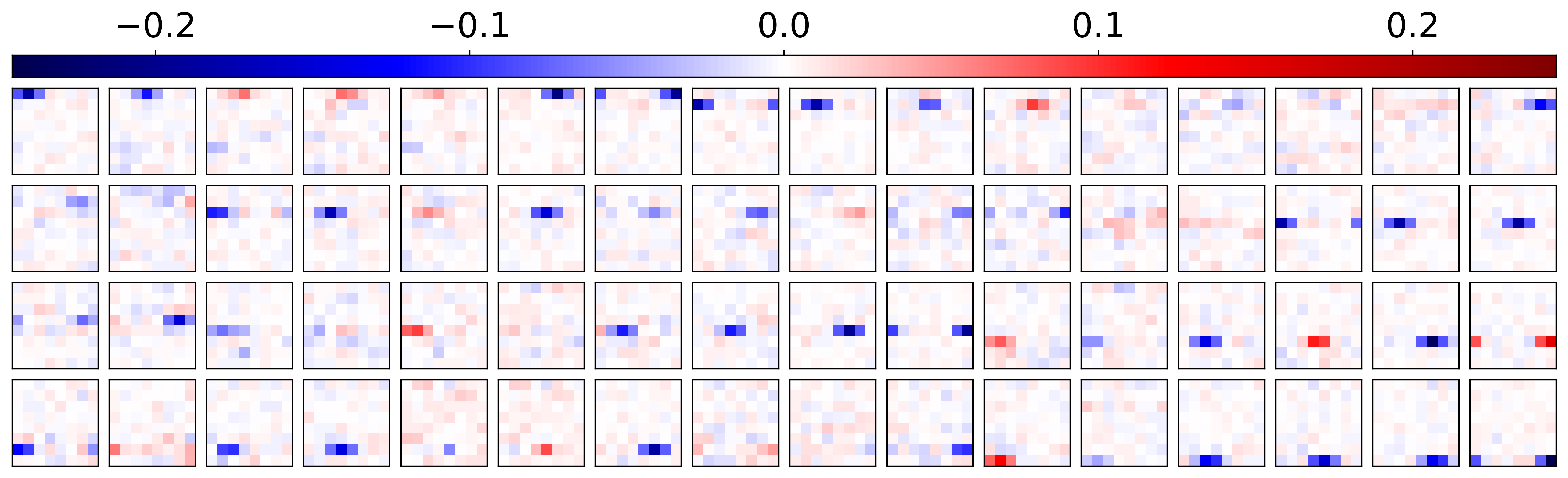}}\\
    \subfloat[]{\label{fig:kitaev_CRBMvsFRBM}\includegraphics[width=0.5\textwidth]{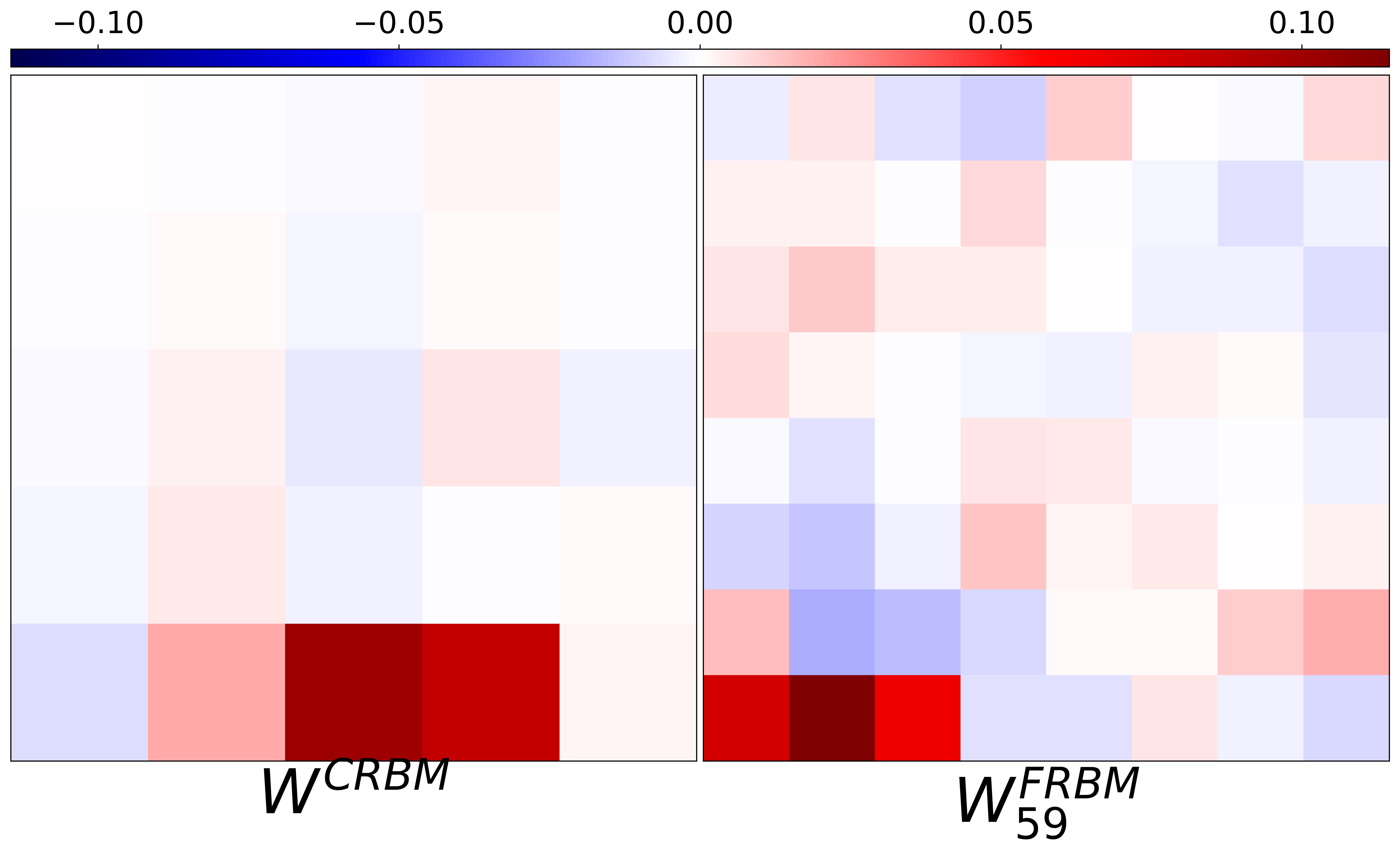}}
    
    \caption{FRBM and CRBM trained at $L=8$ for the Kitaev model at $T=0.018$ with periodic boundary conditions. (a) CRBM with two convolution kernels $W^k$ with size $5\times5$. The lattice points $\eta_r$ interact with an effective interaction with range $4 \times 3$. (b) Weight matrix $W$ of the FRBM, which is reshaped $W_{i,j}\rightarrow W_{i,(j_1, j_2)}$ so that each panel represent an interaction between visible units. The panels were also sorted to show the translational invariance. An effective interaction of $3\times1$ remains. ((c)~left) Kernel of a CRBM trained with one kernel with size $5\times5$. ((c)~right) The 59 component in the figure above.}
    \label{fig:kitaev_kernel}
\end{figure}
\begin{figure*}
    \centering
    \subfloat[]{\label{fig:kitaev_MC_L=8}\includegraphics[width=0.33\textwidth]{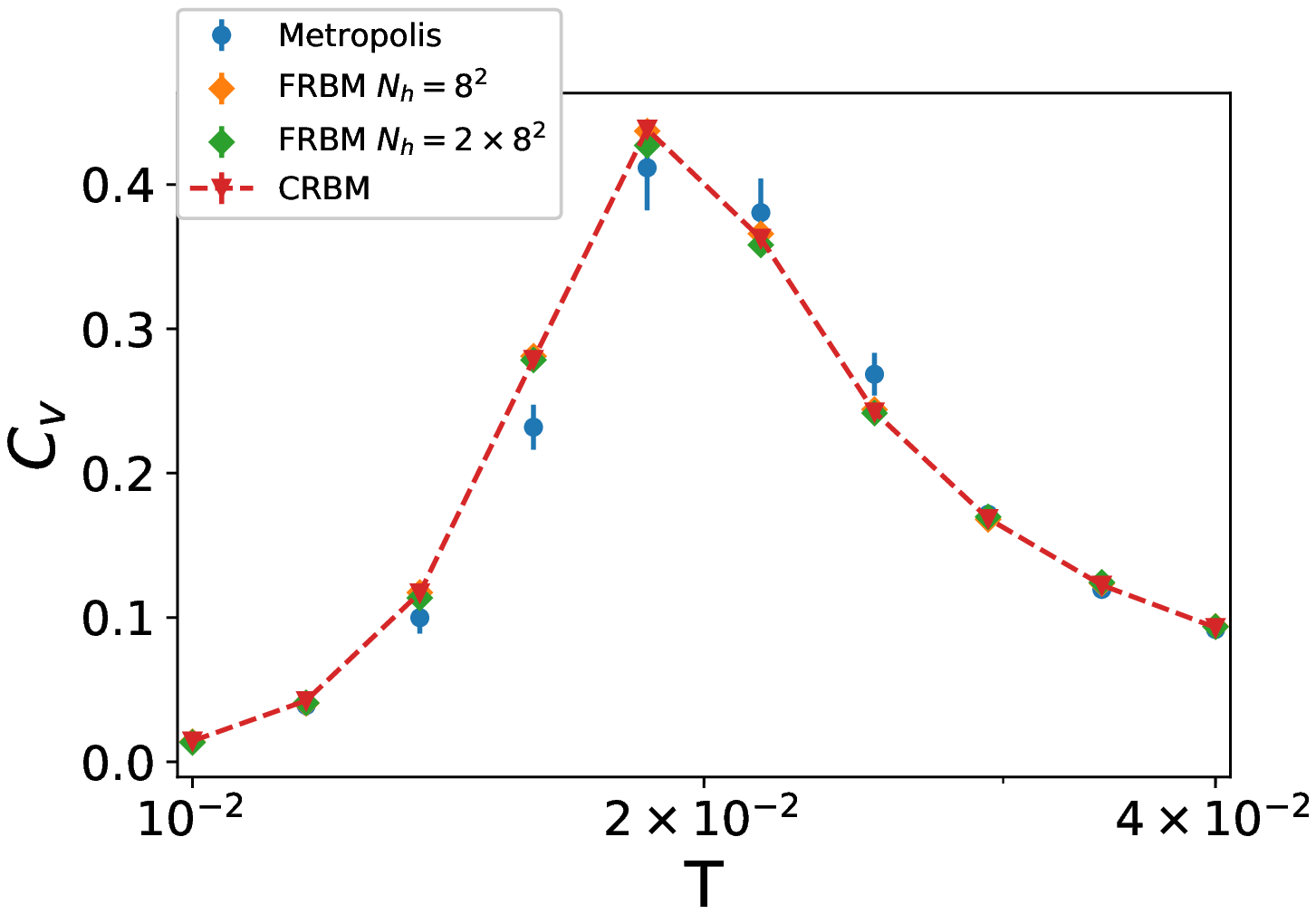}}
    \subfloat[]{\label{fig:kitaev_MC_L=8_tau}\includegraphics[width=0.33\textwidth]{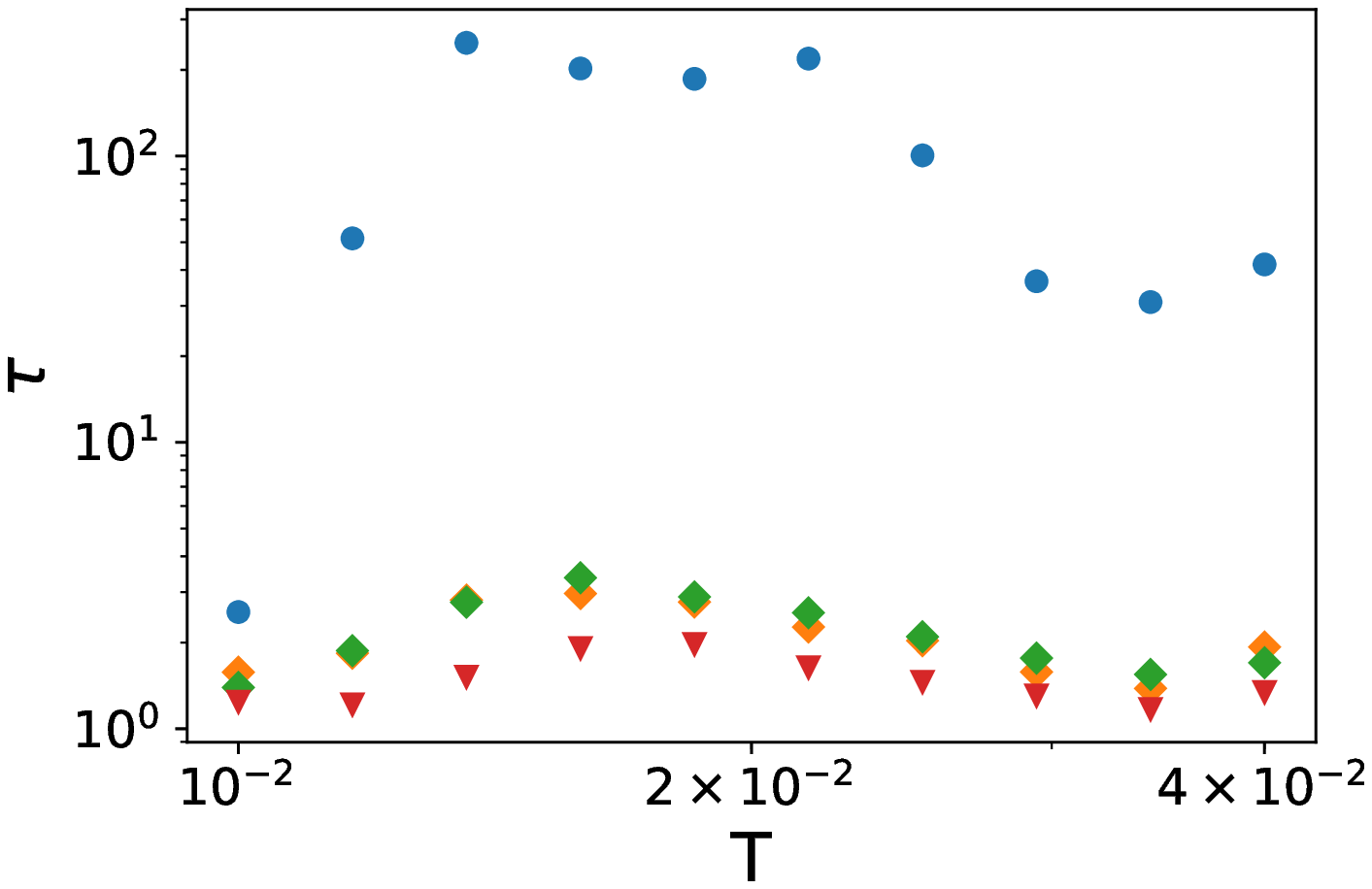}}
    \subfloat[]{\label{fig:kitaev_MC_L=8_accept}\includegraphics[width=0.33\textwidth]{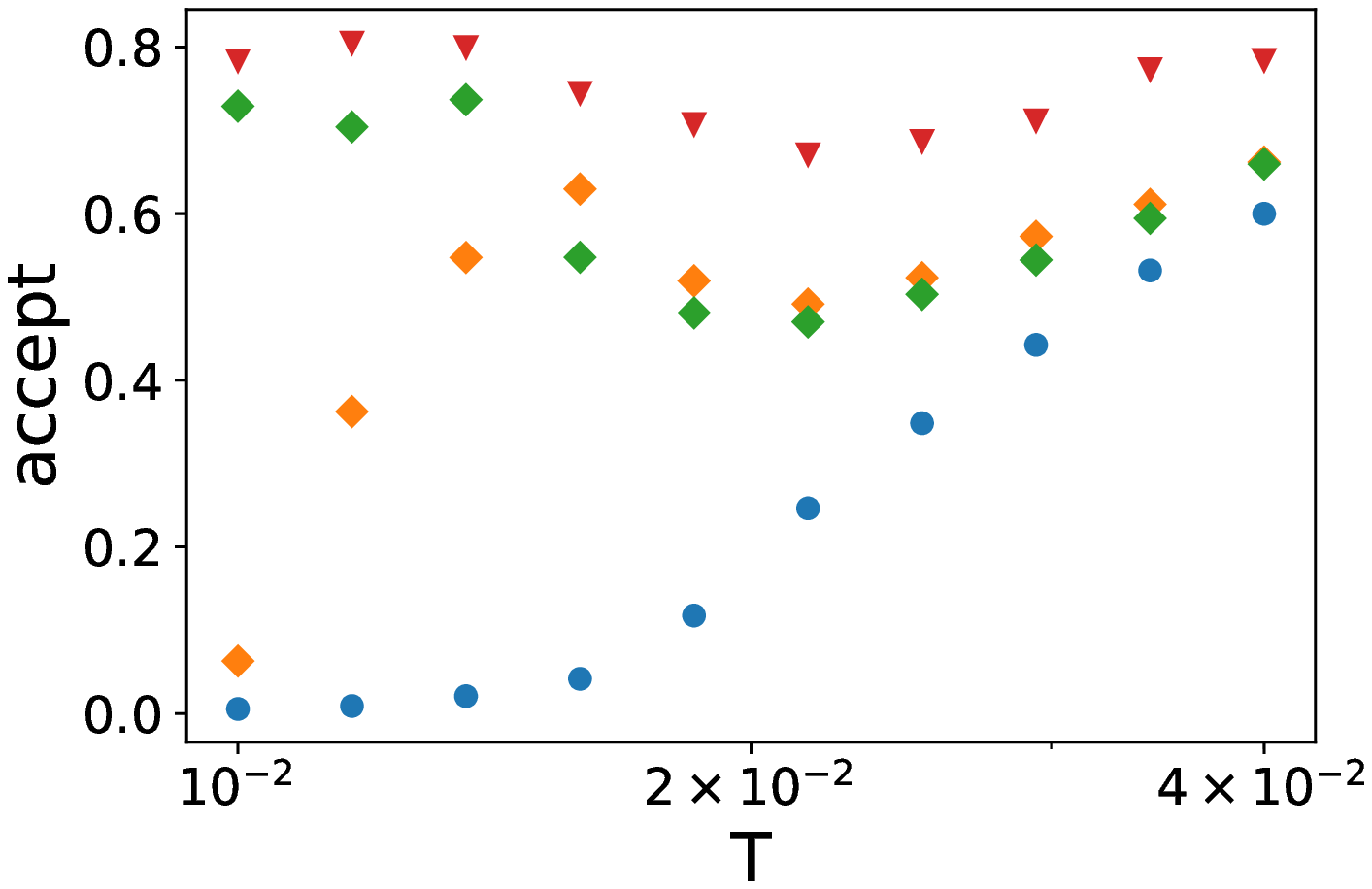}}
    \caption{Numerical results for the Kitaev model at $L=8$ with periodic boundary conditions zoomed around the low temperature crossover generating $4 \times 10^4$ samples. (a) Specific heat, (b) autocorrelation time, and (c) acceptance rate for the Metropolis, FRBM, and CRBM methods.}
\end{figure*}
\begin{figure*}
    \centering
    \subfloat[]{\label{fig:kitaev_MC_N=16}\includegraphics[width=0.33\textwidth]{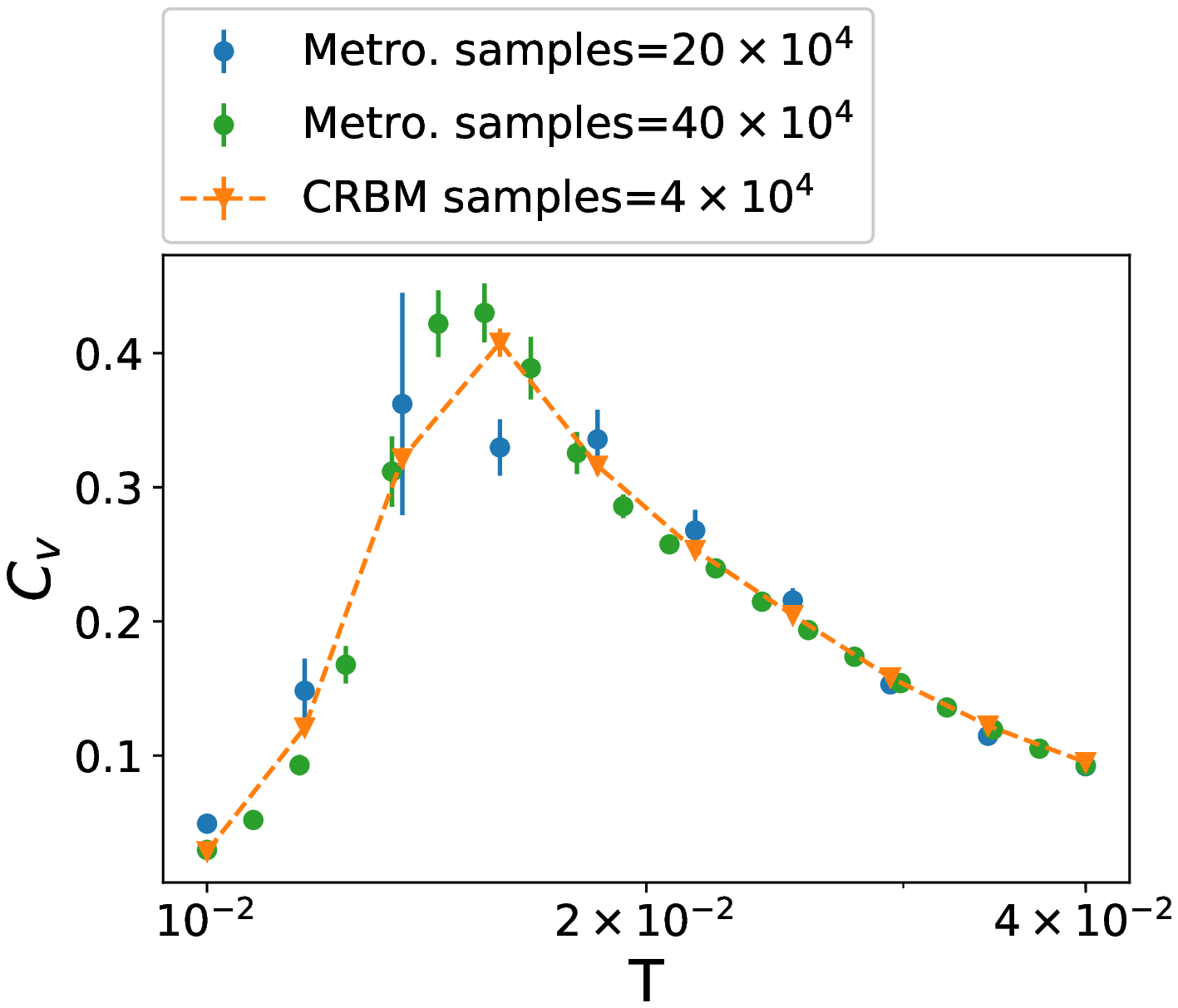}}
    \subfloat[]{\label{fig:kitaev_MC_N=16_auto}\includegraphics[width=0.33\textwidth]{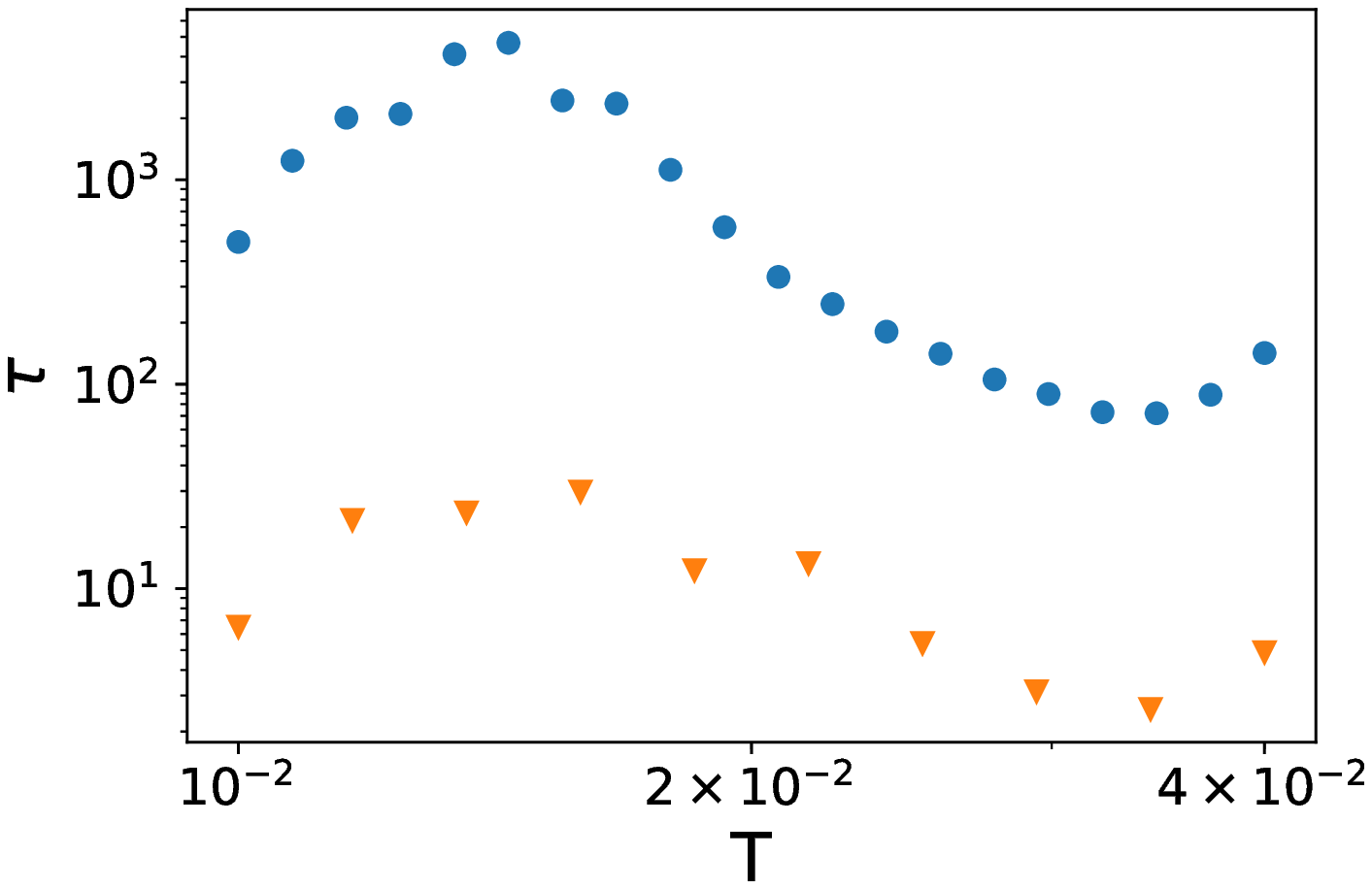}}
    \subfloat[]{\label{fig:kitaev_pretraining}\includegraphics[width=0.33\textwidth]{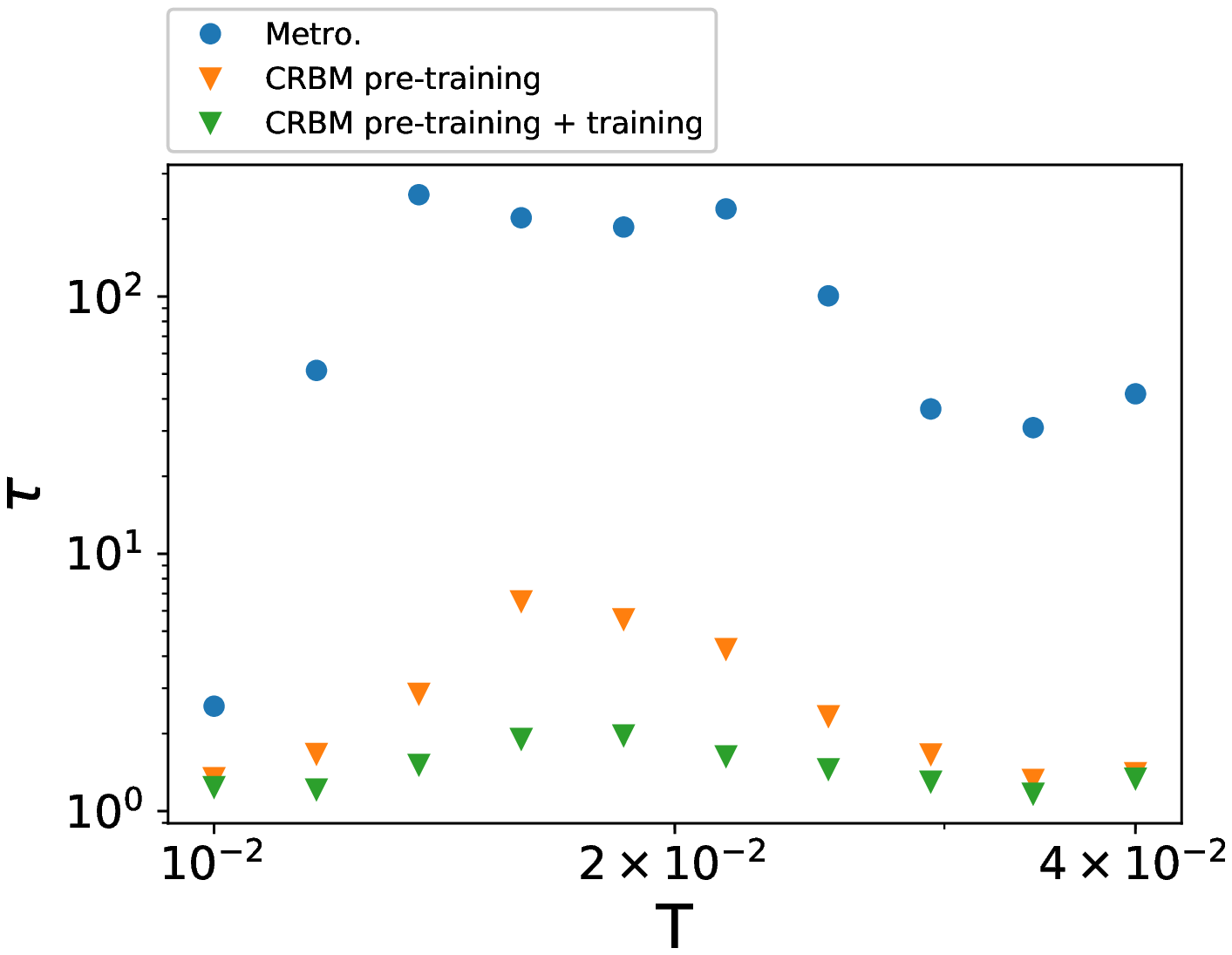}}
    \caption{(a) Specific heat and (b) autocorrelation time at $L=16$ for the Kitaev model with periodic boundary conditions zoomed around the low temperature crossover. (Blue) Metropolis parallel tempering with 10 temperature points. (Orange) CRBM parallel tempering with 10 temperature points. (Green) Metropolis parallel tempering with 20 temperature points. (c) Autocorrelation time with $4\times10^4$ samples at $L=8$. Metropolis is compared to both the CRBM only trained with pre-training and the CRBM completely trained.}
\end{figure*}
In this section, performance will be compared between CRBM, FRBM, and local Metropolis MC with periodic boundary conditions.
First, Metropolis MC simulations using parallel tempering at $L=8$ are performed to generate $4\times10^4$ states at each temperature. Then, for each temperature, a different CRBM with kernel size $2 \times 5\times5$ and an FRBM with a weight matrix with size $8^2\times8^2$ were trained at the lattice size $L=8$.

The trained FRBM and CRBM kernel for $T=0.018$ are plotted in Fig.~\ref{fig:kitaev_kernel}.
Before training the FRBM, all gauge field components can potentially interact, which stands in contrast to the CRBM where only a $5 \times 5$ field of neighboring $\eta_r$ can interact. After training as expected, the trained weight matrix of the FRBM learns to ignore long-range interactions by setting them to 0 and only 3 gauge field interactions in x-y-direction remain. Note that all weight matrix elements in the FRBM are almost the same just shifted around to obtain translation invariance, which is inherent in the CRBM. For the CRBM an effective interaction, where $4\times 3$ field of lattice sites interact with each other, remains after training. 
Interestingly, the FRBM only captures a strong interaction in the x-y-direction but not in the z-direction, while the CRBM shows interactions in both directions. This is still the case if the amount of hidden variables for the FRBM is increased. 

To further understand the origin of the difference between CRBM and FRBM, the CRBM was trained again with a smaller kernel with size $1\times5\times5$. This kernel is compared in Fig.~\ref{fig:kitaev_CRBMvsFRBM} to one of the interaction terms of an FRBM trained with $8^2$ hidden units. When restricting the CRBM to just one kernel it learns the same pattern as the FRBM and accordingly performs similarly. Since the reduced CRBM no longer can capture the full interaction, it focuses on the stronger interaction in the z-direction and neglects the interaction in the x-y-direction. This explains why the FRBM with $8^2$ hidden units performs badly. The FRBM with $2\times8^2$ hidden units should be able to learn the interaction to the same degree as the CRBM since a CRBM with kernel size $2\times5\times5$ can be mapped to an FRBM with $2\times8^2$ hidden units. The question that remains is why in this case the FRBM cannot learn the interaction in the x-y-direction. To find an answer the learned $1\times5\times5$ CRBM kernel is duplicated to yield a kernel of size $2\times5\times5$. When this kernel is trained it does not learn the interaction in the z-direction, as it gets trapped in the local minima where only the interaction in the x-y-direction is learned. We conclude that it is much easier for the FRBM to get stuck in local minima since it not only needs to learn the interaction but also needs to learn to be translationally invariant, which is already encoded in the CRBM.

Next, we analyze the specific heat at different temperatures. We focus only on the low-temperature crossover since the high-temperature crossover does not pose any problem for the MC.
The Metropolis sampling was performed using parallel tempering. For the RBMs, a modified version of parallel tempering is used. Between parallel tempering corrections steps, $k=4$ Gibbs steps were performed with the RBMs (see~App.~\ref{sec:parallel}). Note that $k$ is adapted so that the $k$ Gibbs steps take as much time as one correction step $k=\frac{t_\text{Kitaev}}{t_\text{CRBM}}$ (see~App.~\ref{sec:A_parameters}).

For the lattice size $L=8$, the specific heat values (Fig.~\ref{fig:kitaev_MC_L=8}) for both the CRBM, FRBM, and Metropolis MC agree. Nevertheless, Metropolis MC shows a larger error in the crossover regime. Observe that for the Metropolis algorithm, each update flips one gauge field component at most. In contrast, each sample for the RBMs is almost completely uncorrelated with the previous one. This is also reflected in the integrated autocorrelation time (Fig.~\ref{fig:kitaev_MC_L=8_tau}). The CRBM achieves an autocorrelation time that is 100 times smaller than that by Metropolis MC. The decreased autocorrelation will make it possible to sample effectively at large system sizes. The CRBM autocorrelation is also two times smaller than the one achieved by the FRBM. This is still the case if the amount of hidden units is increased by a factor of two. The CRBM has smaller autocorrelation than the FRBM due to a smaller loss during training, which can be traced back to the inability of the FRBM to learn the interactions in the z-direction. This is also reflected in the acceptance rate, which is larger for the CRBM. One can also see that the FRBM with more hidden units achieves a larger acceptance rate than the other FRBM at low temperatures.

For $L=16$, Metropolis acquires problems in converging at temperatures close to the crossover region as can be seen in Fig.~\ref{fig:kitaev_MC_N=16}. The temperature points have to be increased by a factor of two, to take advantage of parallel tempering so that the results are more stable and also increase the sample size to $40\times10^4$. In contrast, the CRBM works well with 10 times fewer samples and half as many temperature points, even though it was trained at $L=8$. This is explained by the autocorrelation time, which close to the crossover region is about 150 times larger for Metropolis. Note that the use of a fully connected RBM would not be possible here due to the increasing need for samples due to the larger lattice size. Besides, it would need to learn from the poorly performing Metropolis.
\subsubsection{Pre-training}
\begin{figure*}
    \subfloat[]{\label{fig:kitaev_mc_N=16_noboundry}\includegraphics[width=0.33\textwidth]{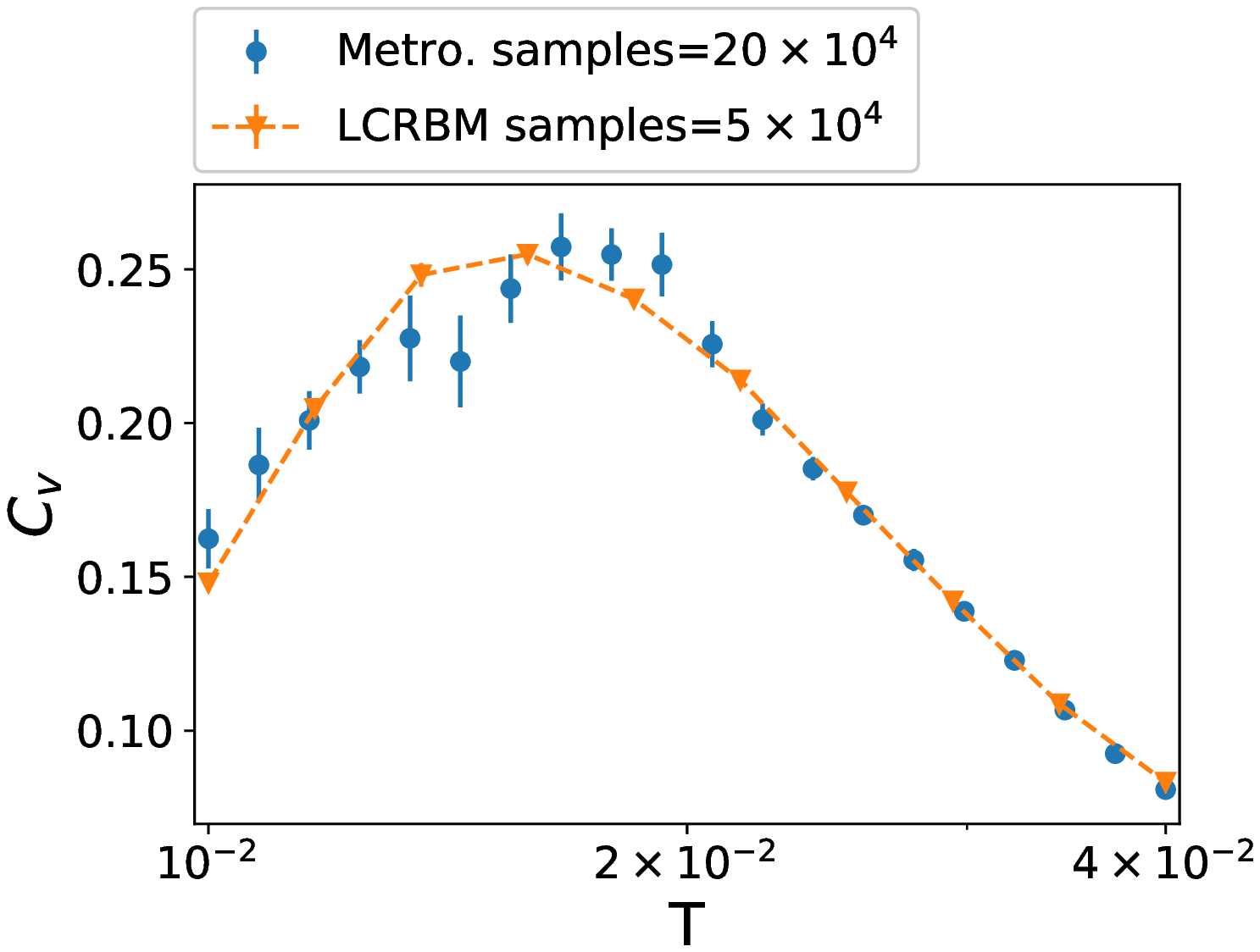}}
    \subfloat[]{\label{fig:kitaev_accept_N=16_noboundry}\includegraphics[width=0.33\textwidth]{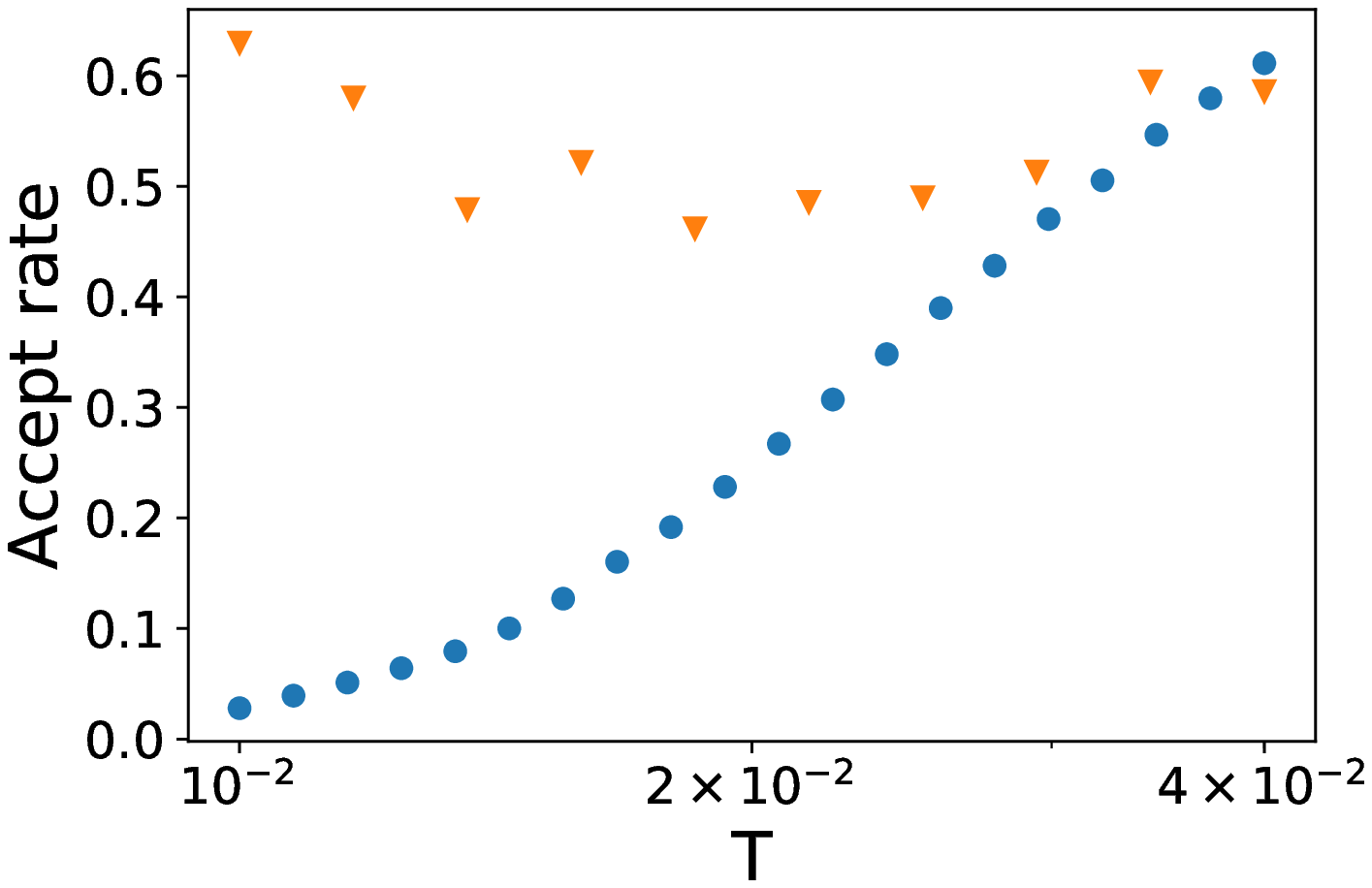}}
    \subfloat[]{\label{fig:kitaev_auto_N=16_noboundry}\includegraphics[width=0.33\textwidth]{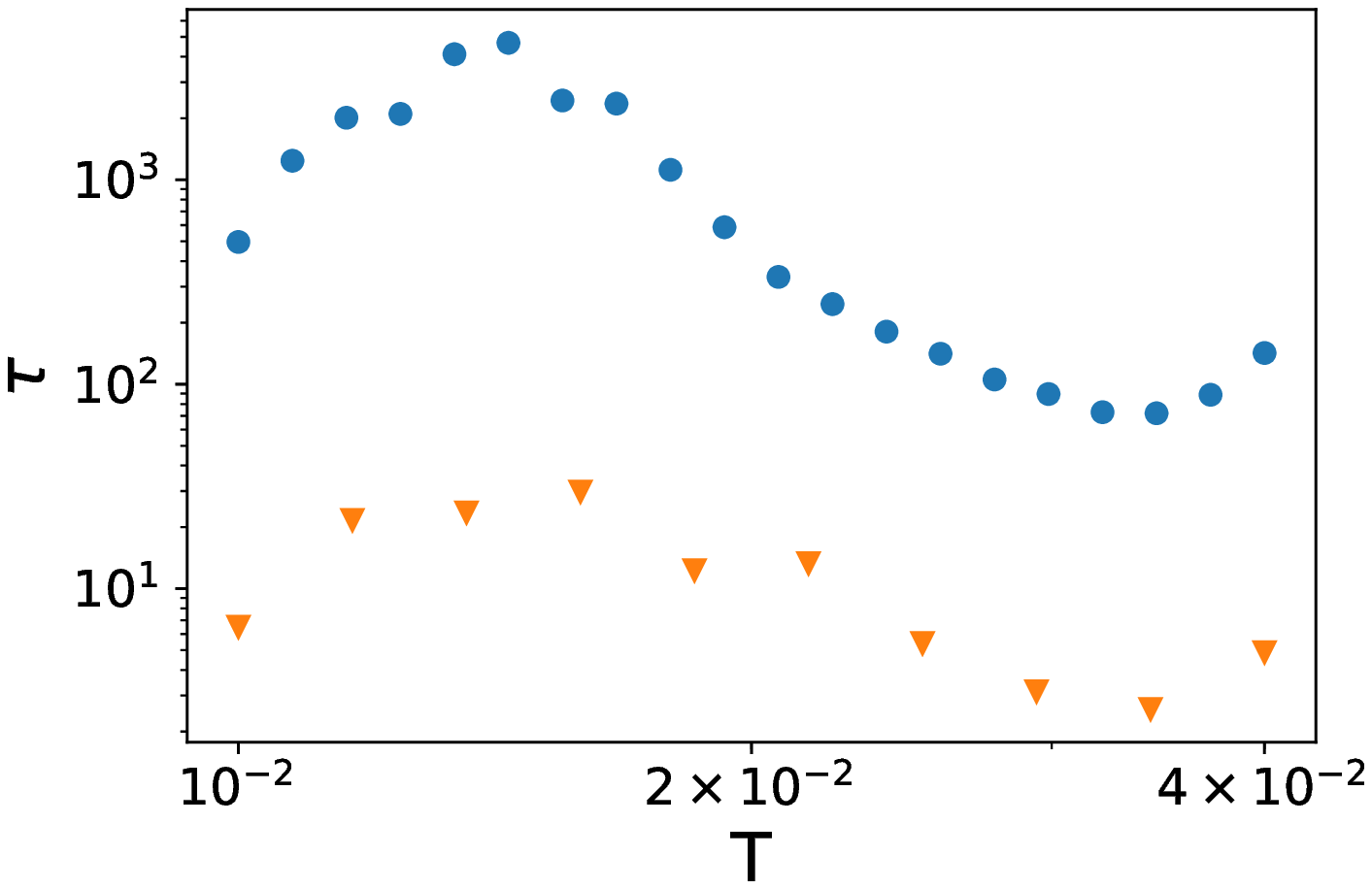}}
    \caption{Numerical results for the Kitaev model with $\alpha=1$, $L=16$, and open boundary conditions for the (a) specific heat,  (b) acceptance rate of the parallel tempering exchange between the physical distribution and the LCRBM, and (c) autocorrelation time of the energy. Blue points refer to the Metropolis parallel tempering with 20 temperature points and orange triangles are CRBM parallel tempering with 10 temperature points. }
\end{figure*}
We discuss the effect of the pre-training step where the CRBM is first trained with $10^4$ states generated at random.
The cheap pre-training mainly speeds up the convergence of training. To train the CRBM without the pre-training step, the model needs to be trained on average 200 epochs to reach the minimum of the loss. In contrast, this is reduced to an average of 30 epochs if pre-training is applied beforehand. Not only that but if only the pre-training is done, the CRBM still performs better then the Metropolis algorithm, see Fig.~\ref{fig:kitaev_pretraining}. Note that in order to perform the pre-training step, no Metropolis MC simulation needs to be carried out beforehand. This gives us the possibility to sample the physical states from the pre-trained CRBM and then use those same states for finalizing its training. This was tested and gives similar results to training with states sampled with local Metropolis.
\subsection{Open Boundary Conditions}
\begin{figure*}
    \centering
\subfloat[]{\label{fig:kitaev_mc_N=24_noboundry}\includegraphics[width=0.33\textwidth]{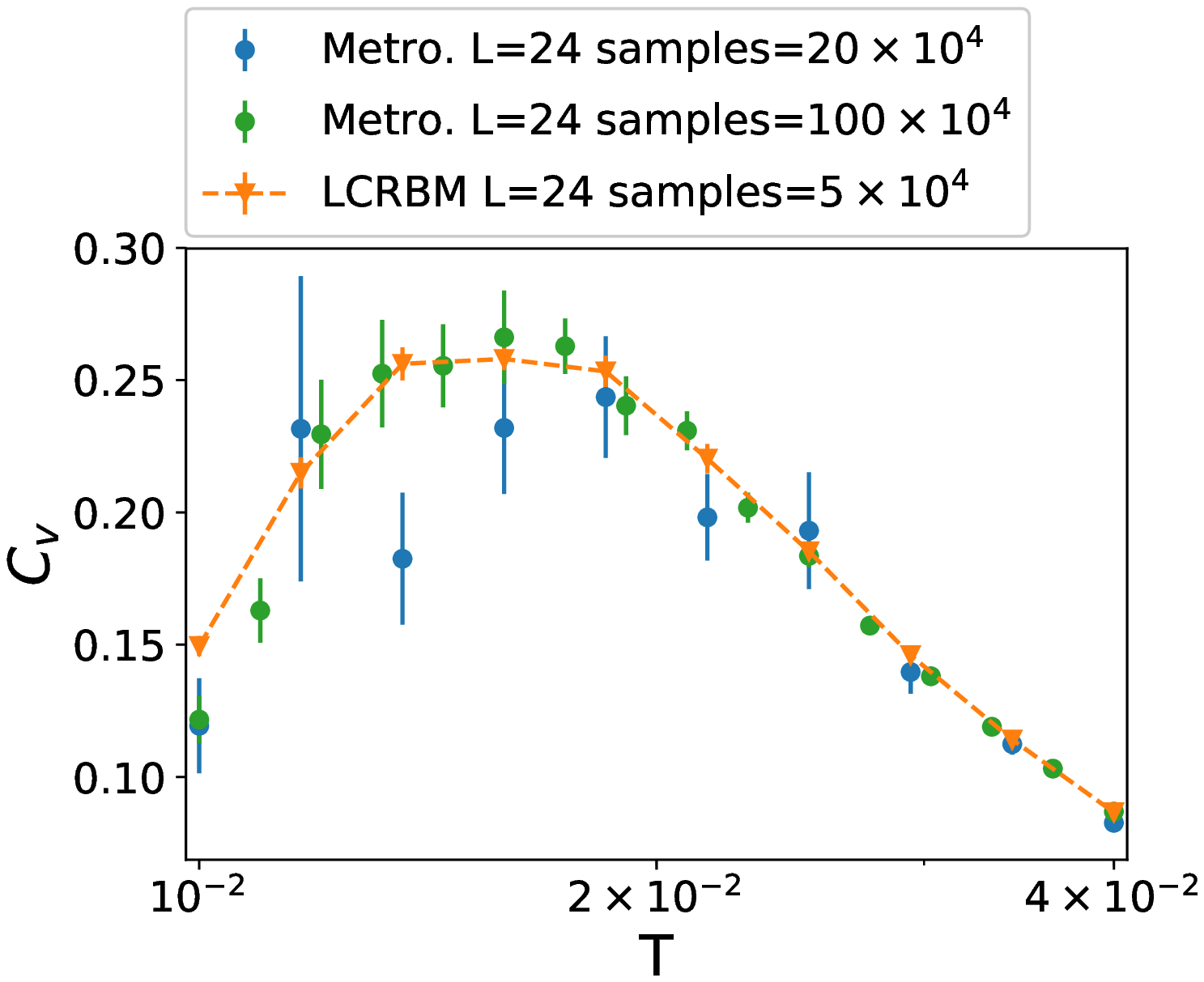}}
\subfloat[]{\label{fig:kitaev_energies_N=24_noboundry}\includegraphics[width=0.33\textwidth]{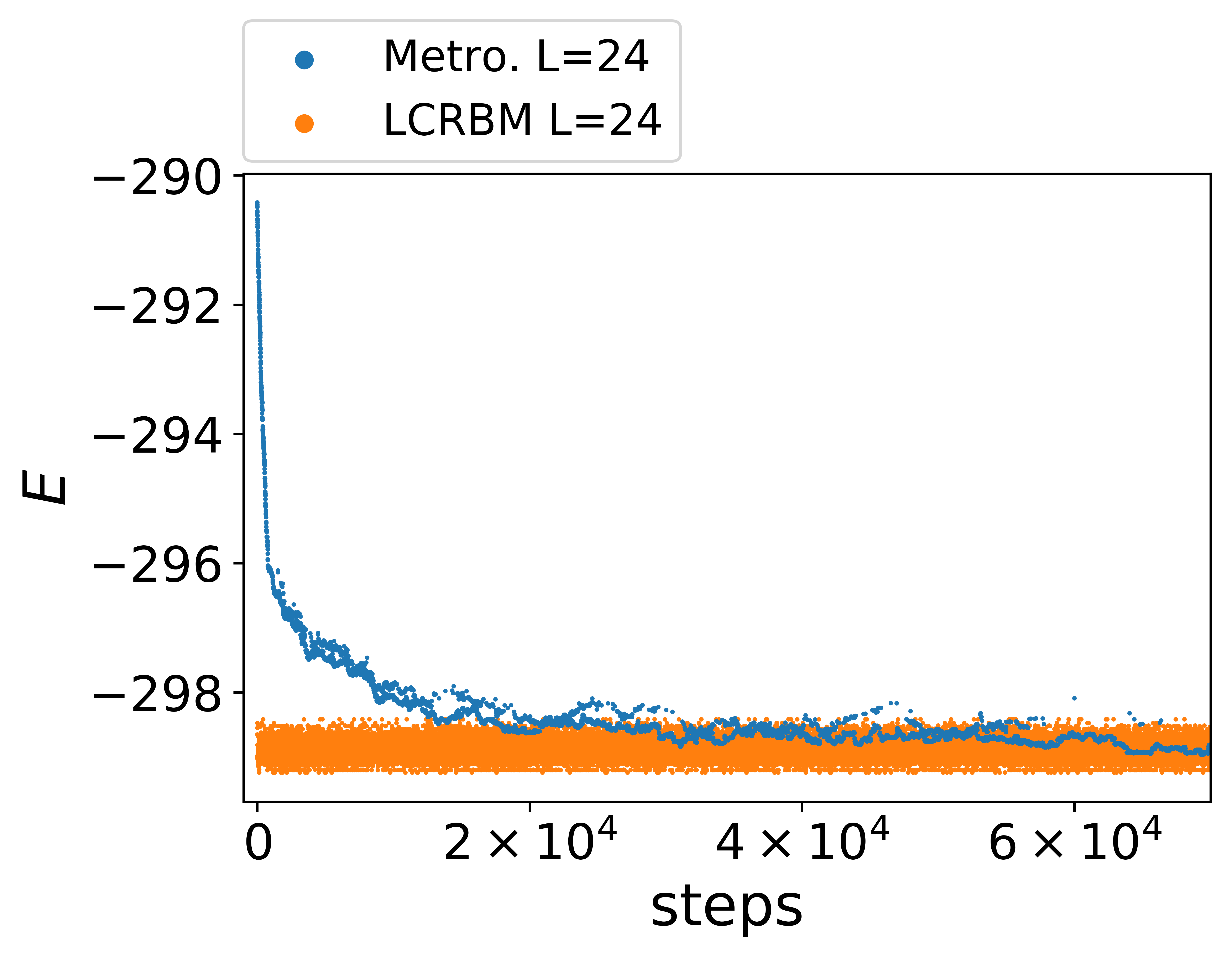}}
\subfloat[]{\label{fig:kitaev_mc_N=12-24_noboundry}\includegraphics[width=0.33\textwidth]{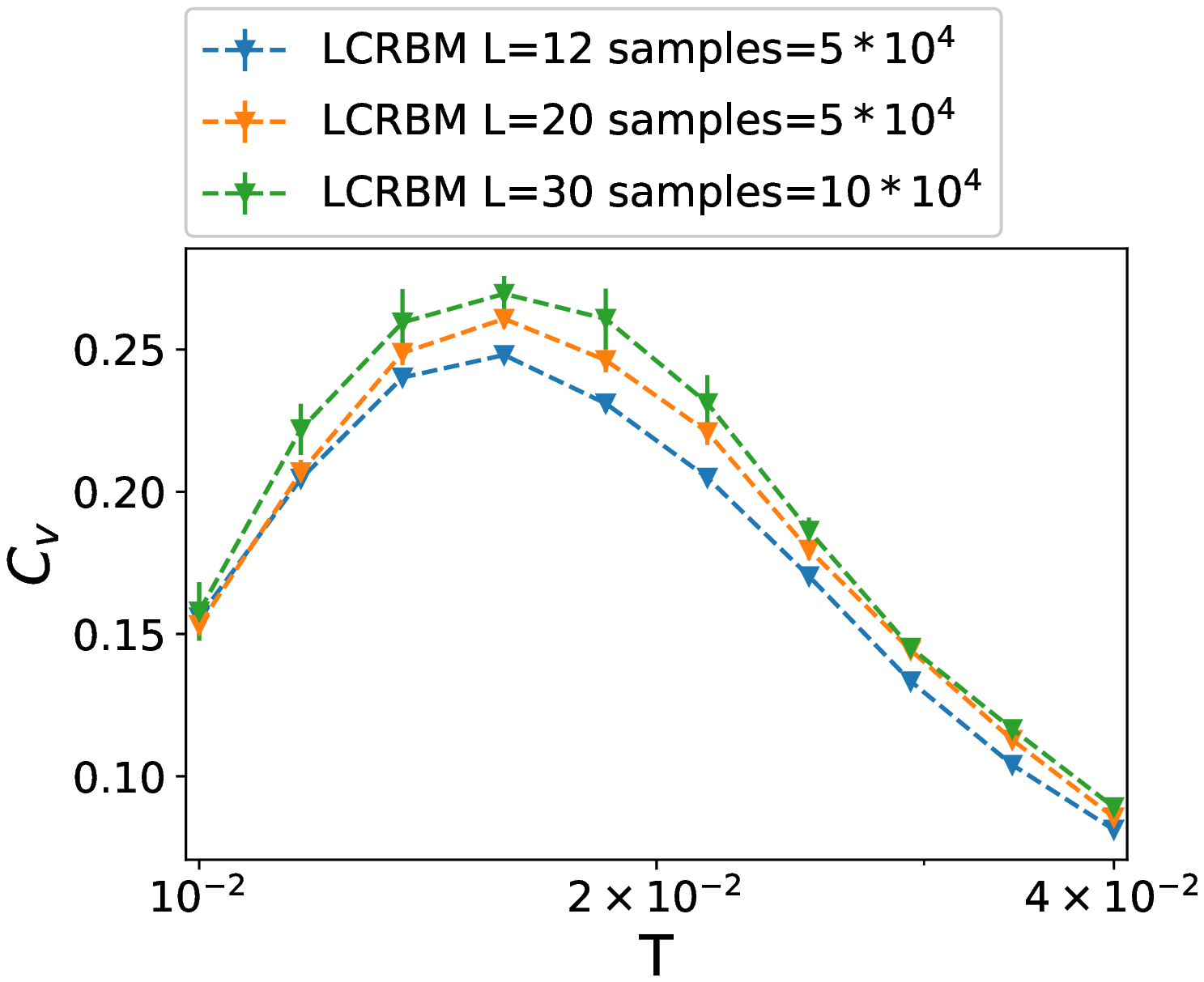}}
\caption{(a) Specific heat around the first low temperature crossover with $\alpha=1$ and $L=24$ at different temperatures with open boundary conditions. (b) Energies of samples during the MC, for Metropolis and the LCRBM, at $T=10^{-2}$. Panel (c) compares specific heat obtained with the LCRBM at $L=\{12,20,30\}$. Note that four times more samples were generated for the Metropolis algorithm in order to reduce errors to obtain agreement between the two methods.}
\end{figure*}
To treat the open boundary conditions we adapt the CRBM by using the locally connected CRBM (LCRBM) to learn the free energy of a system with open boundary conditions in x-y-direction and periodic boundary conditions in the z-direction. The LCRBM is similar to the CRBM in that in the bulk of the lattice a  translation invariance is assumed, i.e. the weights are shared. The two models are, however, different in that for the LCRBM new unshared kernels are introduced at the edges. For this purpose, three unshared kernels are introduced on the left and right sides to learn boundary effects. Instead of training a new LCRBM, we use the values of the CRBM used for periodic boundaries and then retrain it at $L=12$, which converges after only a few epochs. The kernels before training and after are very similar. For $L=16$, the specific heat is compared between LCRBM and Metropolis (see~Fig~\ref{fig:kitaev_mc_N=16_noboundry}). The LCRBM has an extremely low error even though 8 times fewer computations where performed, which is explained by the lower autocorrelation time. This is especially noticeable at lower temperatures and close to the crossover.

Another aspect of interest is the acceptance rate and the autocorrelation time, Fig.~\ref{fig:kitaev_accept_N=16_noboundry},c.
The acceptance rate for Metropolis increases from almost zero to $0.6$ with increasing temperatures. This is expected as smaller $T$ translate directly to smaller acceptance rates. For the LCRBM the acceptance rate is not strongly temperature-dependent, as it is mostly influenced by how well the LCRBM fits the physical probability distribution. One has to keep in mind that each time a state is accepted by Metropolis a maximum of one gauge component has been flipped. In contrast, each state accepted by the physical distribution from the LCRBM is almost independent of its predecessor. This is confirmed by the autocorrelation time.  The autocorrelation of the LCRBM at the crossover is an average of $\sim 100$ times smaller than Metropolis.
Taking into account that more temperature points are needed, this is equivalent to a $\sim 200$ times faster simulation. Note that to obtain error values that are small enough using Metropolis, four times more samples had to be generated.

Increasing the lattice size to $L=24$, the Metropolis method does not converge, whereas the LCRBM trained at $L=8$ converges easily (Fig.~\ref{fig:kitaev_mc_N=24_noboundry}). For Metropolis the warm-up phase had to be increased to $20\times10^4$ and $100\times10^4$ states had to be sampled. This can be traced back to the high autocorrelation time. In Fig.~\ref{fig:kitaev_energies_N=24_noboundry} the energies during the MC at low a temperature are shown. The Metropolis method takes a long time until equilibrium is reached and one can see strong correlations between samples. In contrast, the LCRBM is in equilibrium after the first LCRBM parallel tempering step, as one LCRBM parallel tempering step consists of $140$ Gibbs steps ($L=24$) and one corrections step.

The specific heat for $L=12,20,30$ as shown in Fig.~\ref{fig:kitaev_mc_N=12-24_noboundry} only show small differences between each other in agreement with Motome \etal~\cite{2Dkitaev}. For 2D there is no phase transition just a crossover since the specific heat is not singular at $L \rightarrow \infty$.

%% file: conclusion.tex
To conclude we employ convolutional restricted Boltzmann machines to learn an effective energy for the 2D Ising and Kitaev models. In contrast to the fully connected RBMs, which suffer from long training times when the lattice size is increased, CRBM was shown to be more efficient because it can be trained for smaller lattice sizes before applying to the larger lattices using translation invariance of the model. It was also shown that for the Kitaev model the CRBM better captures the physical interaction than the FRBM since it does not need to learn to be translational invariant. We showed that, not only can a CRBM reproduce thermodynamic observables accurately and give results with smaller errors for both periodic and open boundary conditions for the Kitaev model, but that it is also able to simulate lattice sizes up to $L=30$, which is not possible using the local Metropolis algorithm in any reasonable time. Our results also confirm that the Kitaev model does not possess a phase transition but only a crossover. It is of interest to explore in future Kitaev-like model systems with a phase transition, as the CRBM could enable more accurate finite-size scaling.

\section{Acknowledgment} 

We thank N.B. Perkins for the stimulating discussions.

%% file: Atraining.tex
\label{sec:Atraining}
The physical model with energy $E_\text{phys}(x)$ will be approximated by the RBM. States in the physical model are encoded as, $-1$ and $+1$. The RBM encodes its states as, $0$ and $1$. While sampling and training, a conversion between the two is necessary.

Training is done supervised similarly to Ref.~\cite{huang2017accelerated} by minimizing the loss function:
\begin{align}
\text{loss}(W) = \frac{1}{M}\sum_{i=1}^M [E_\text{phys}(x_i)\beta - F_\text{RBM}(x_i;W) - C(W)]^2 
\end{align}
where $L$ is the lattice size, $\beta$ is the inverse temperature and $C$ is a value that can be chosen freely because the probability function is invariant under the addition of a constant to the energy. 
$C$ is chosen such that the loss is minimal:
\begin{align}
C(W) = \frac{1}{M} \sum_{i=1}^M E_\text{phys}(x_i)\beta - F_\text{RBM}(x_i;W).
\end{align}
In practice, to minimize the loss ADAM batch-gradient descent is used. A combination of samples from both the physical distribution $P_\text{phys}$ and the RBM's distribution $P_\text{RBM}$ are used for training. These are expensive, as after each training step $P_\text{RBM}$ changes. This means that new samples need to be generated for training. In addition, in each training step, $E_\text{phys}(x)$ for these new samples has to be computed, which is numerically expensive.
To alleviate the computational load, we perform two additional steps. First, a pre-training step is performed in which completely random states are generated and trained until the loss of the physical states no longer decreases. This reduces the training time because $E_\text{phys}(x)$ only needs to be computed once for each random state, and the random states can be used to pre-train the RBM at all temperatures. The second step is to use a buffer of past RBM samples where instead of generating new samples in each training step, only a portion of the used states are new and the rest are reused from previous training steps. In each epoch, $ub$ new samples are added to the buffer. In practice, a value of $ub=200$ is chosen.
Below we give the details of the algorithms involved.
\begin{algorithm}[H]
		\caption{Train the RBM}
		\label{alg:train}
		\begin{algorithmic}
			
		\State $N \gets$ train size
		\State $M \gets$ batch size
		\State $K\gets$ amount of used kernels
		\State $L_W\gets$ kernel size
		\State $ub\gets$ Buffer update size
		\State initialize $W \sim \mathcal{N}\left(m=0,\sigma=\frac{2}{KL_W^2}\right)$ with size $K \times L_W\times L_W$
		\State initialize $v_\text{bias}=0$ with size $1$
		\State initialize $h_\text{bias}=0$ with size $K$
		\State $param=[W, v_\text{bias}, h_\text{bias}]$
		\State sample $x_\text{phys} \sim P_\text{phys}$ with Metropolis
		\State calculate $E_\text{phys}(x_\text{phys})$
		
		\State initialize $\bar{x}_\text{RBM} \sim \text{Binomial(size}=(ub, L, L), p=0.5)$
		\State
		\Function {update buffer}{}
		
		\State update $\bar{x}_\text{RBM}$ with $k$ steps of Gibbs sampling
		\State $x_\text{RBM} \gets x_\text{RBM} \cup \bar{x}_\text{RBM}$
		\State calculate $E_\text{phys}(\bar{x}_\text{RBM})$
		\If{size($x_\text{RBM}$) $> N$}
		\State remove last elements of $x_\text{RBM}$
		\EndIf
		\EndFunction
		\State
		\State iterate UPDATE BUFFER until size($x_\text{RBM}$)$=N$
		\State
		\While{loss $> \epsilon$}
		\State UPDATE BUFFER 
		
		\State $x_{tot} = x_\text{RBM} \cup x_\text{phys}$
		
		\For{x = ($M$ elements of $x_{tot}$)}
		
		\State $\text{diff}^i = E_\text{phys}(x^{(i)})\beta - F_\text{RBM}(x^{(i)};param)$
		\State $C = \frac{1}{M}\sum_{i=1}^M \text{diff}^i$
		\State loss = $\frac{1}{M}\sum_{i=1}^M (\text{diff}^i - C)^2$
		\State $param \gets param - \text{ADAM}\left(\frac{\partial \text{loss}}{\partial param}\right)$
		
		\EndFor
		
		\EndWhile\label{euclidendwhile}
		
		\end{algorithmic}
	\end{algorithm}

%% file: Asampling.tex
\subsection{Gibbs sampling}
\label{sec:gibbs}
The standard way to sample from an RBM is Gibbs sampling. The conditional probability $P(v|h)=\frac{P(v,h)}{P(h)}$ is computed to:
\begin{align}
P(h_i = 1| v) &= \frac{e^{h_\text{bias}^i +\sum_j W_{ij}v_j}}{1 + e^{h_\text{bias}^i +\sum_j W_{ij}v_j}},\\
 &= \sigma(h_\text{bias}^i +\sum_j W_{ij}v^j),
 \\
 P(v_j = 1| h)&= \sigma(v_\text{bias}^j +\sum_i h_i W_{ij}).
\end{align}

The conditional probabilities describe a binomial distribution that depends either on $h$ or on $v$. A bipartite Markov chain is constructed. First, a sample $h^{(1)}$ is drawn from the conditional distribution $P(h^{(1)}|v^{(1)})$. Second, the new visible sample is drawn from $P(v^{(2)}|h^{(1)})$.
\begin{align*}
v^1 \xrightarrow{\LARGE \displaystyle P(h^1|v^1)} h^1 \xrightarrow{ \LARGE \displaystyle P(v^2|h^1)} v^2.
\end{align*}
The only difference between this and a normal Markov chain is the middle step where $h$ is sampled. The RBM can be reformulated so that it is equivalent to a normal Markov chain, as the transition probability from $v^{(1)}$ to $v^{(2)}$ is:
\begin{align}
P(v^{(2)}|v^{(1)})=\sum_h P(v^{(2)}|h)P(h|v^{(1)}).
\end{align}
Note that this quantity can not be computed analytically, but it can be proven that this conditional probability fulfills the detailed balance equation. This means that the RBM describes an ergodic Markov chain.
\subsection{Parallel tempering}
\label{sec:parallel}
\begin{figure}
	\centering
	\subfloat[]{\label{fig:parallel_metro_tempering}\includegraphics[width=0.45\textwidth,valign=t]{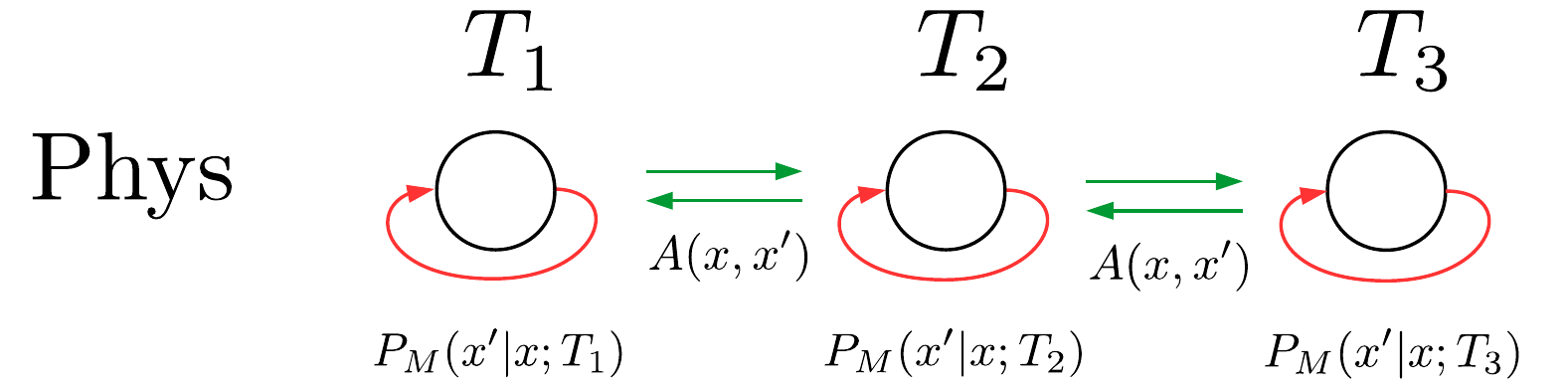}}
	 \hfill 
	\subfloat[]{\label{fig:parallel_crbm_tempering}\includegraphics[width=0.45\textwidth,valign=t]{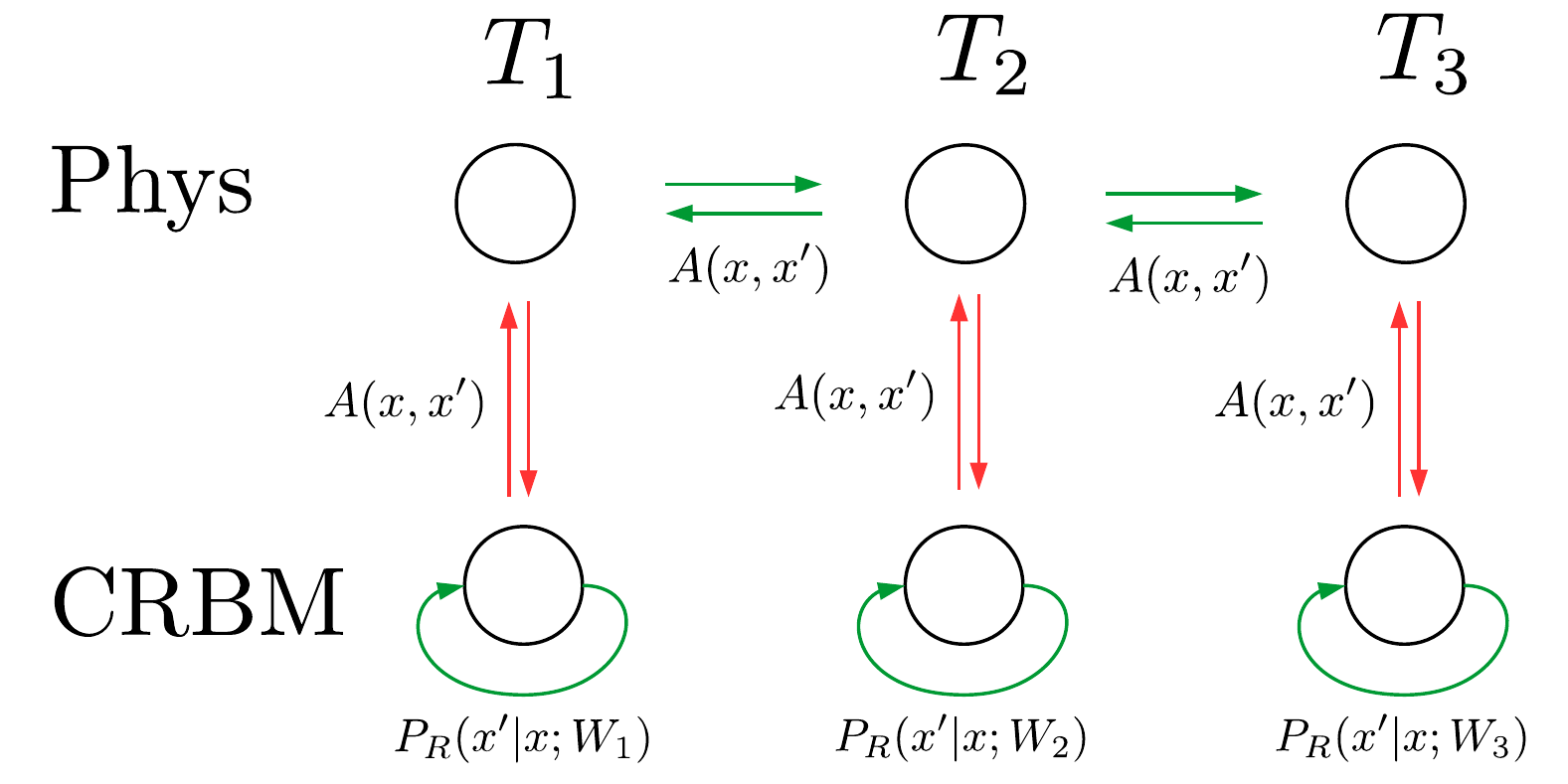}}
	\caption{(a) Parallel tempering step where $P_M(x'|x)$ represents a local Metropolis MC step and $A(x',x)$ (Eq.~\ref{eq:acceptance}) is the probability that the two temperatures exchange states. Green arrows represent a cheap operation. Red arrows represent an expensive operation. (b) Parallel tempering exchange correction where instead of updating with Metropolis states, exchanges with CRBMs are introduced. A $P_R$ step represents $k$ Gibbs steps.}
\end{figure}

Often problems plague Metropolis MC simulations at small temperatures because the acceptance rate is very low. A method that tackles this is parallel tempering.
Parallel tempering takes advantage of the fact that probability distributions at similar temperatures overlap. It introduces an exchange between high and low temperatures so that the low-temperature dynamics can take advantage of the high acceptance rates at high temperatures.
States are exchanged at different temperatures with Metropolis acceptance probability:
\begin{align}
\label{eq:acceptance}
A(x_1,x_2) = \min( 1, e^{F_1(x_1)\beta_1 + F_2(x_2)\beta_2-F_1(x_2)\beta_1-F_2(x_1)\beta_2})
\end{align}
For the Kitaev model, exchanges are computationally cheap because the eigenvalue problem for both states has already been solved. In Fig.~\ref{fig:parallel_metro_tempering}, the full sampling cycle can be seen. First, a Metropolis step and then a state exchange with the neighboring temperature is performed. 

\begin{figure}
    \centering
    \subfloat[]{\label{fig:kitaev_MC_L=8_nocorr}\includegraphics[width=0.4\textwidth]{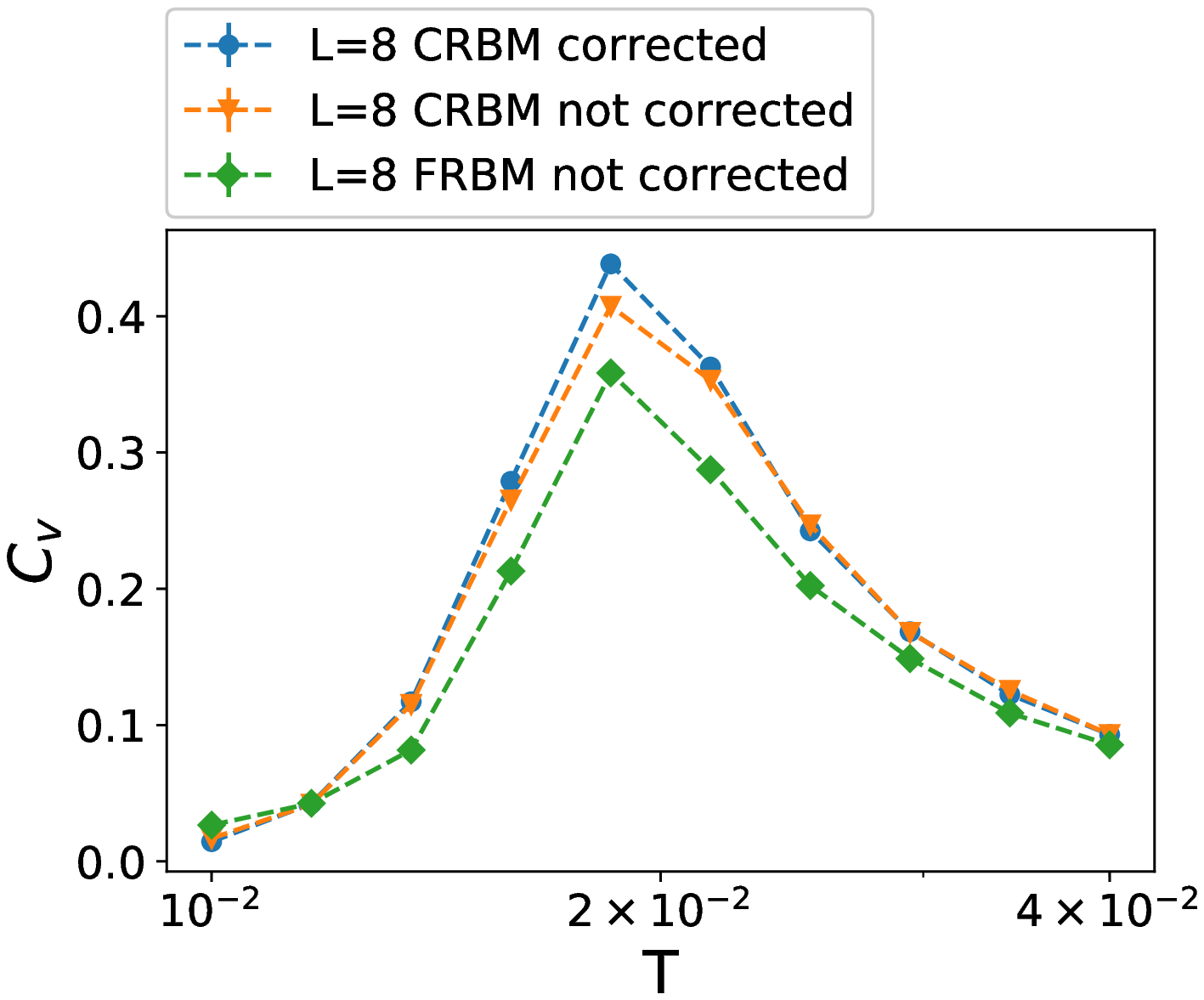}}
    \subfloat[]{\label{fig:kitaev_MC_L=24_nocorr}\includegraphics[width=0.4\textwidth]{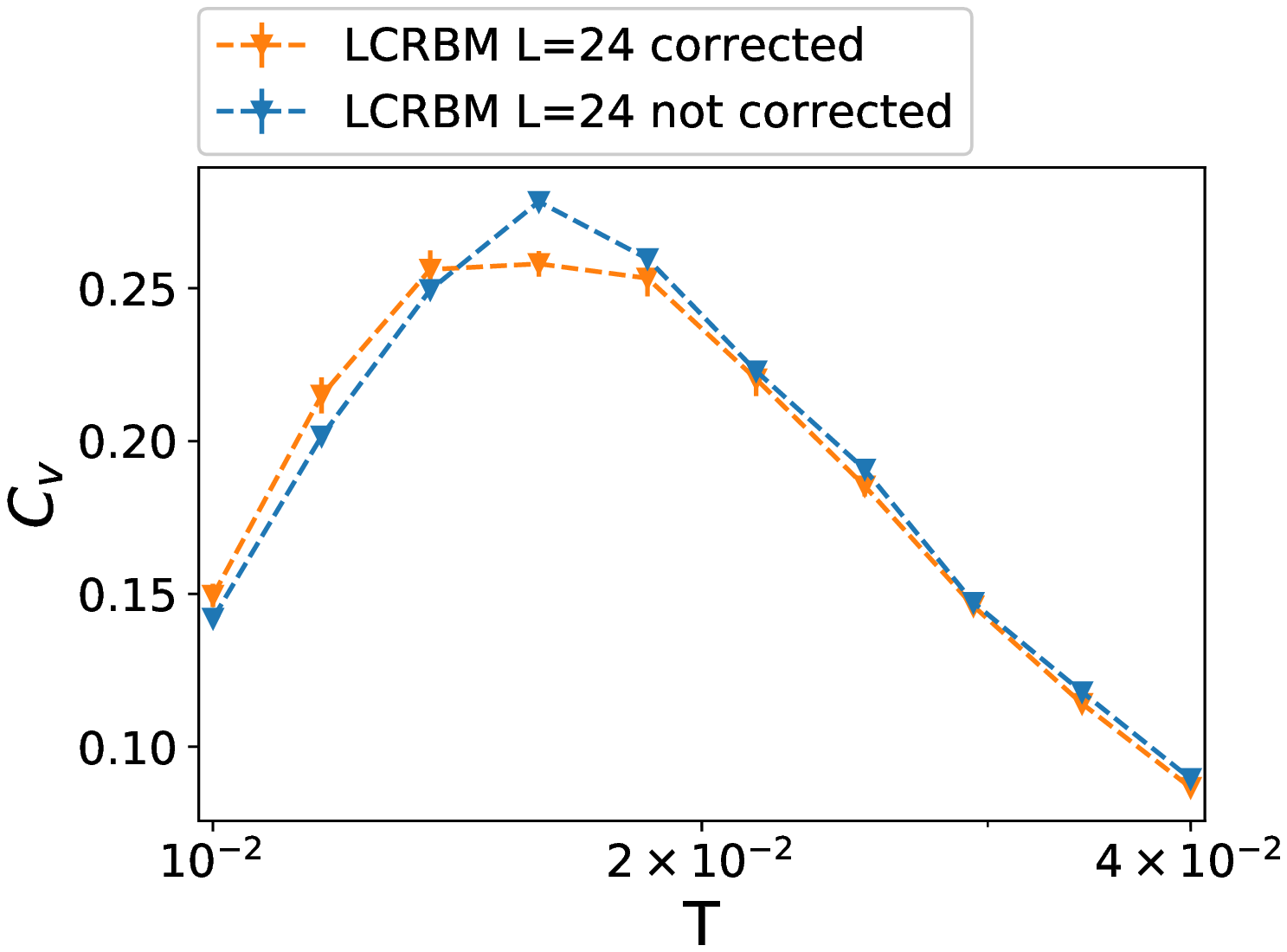}}
    \caption{Specific heat results compared between various RBMs that where statistically corrected with the parallel tempering exchange correction and ones without correction procedure. Here, $4\times10^4$ samples were generated. Panel (a) shows a comparison of FRBM and CRBM at $L=8$ with periodic boundary conditions, and (b) refers to LCRBM at $L=16$ with open boundary conditions.}
\end{figure}

A small modification leads to parallel tempering with an exchange correction. Instead of using Metropolis MC for the updates, we introduce the RBM as a different probability distribution that the physical states can exchange with. The more the approximate RBM distribution and physical distribution overlap, the larger the acceptance rate is (see~Eq.~\ref{eq:acceptance}). Sampling entails three steps as seen in Fig.~\ref{fig:parallel_crbm_tempering}. First, the RBM states are updated through Gibbs sampling a certain amount of times (in practice we use $60$ steps). The RBM states are then exchanged with the physical distribution with probability $A(x_1,x_2)$. Lastly, a normal parallel tempering exchange is performed. Each time a state is accepted, it is almost uncorrelated to the previous one. This is in stark contrast with local Metropolis MC simulations where only one lattice point is updated each time as a new state is accepted. Note that physical exchanges are cheap because one does not need to recalculate the eigenvalue problem to know what the free energy at a different temperature is. Note that this method was not used for the Ising model.
\\ \\
It is interesting to see what would happen if the RBM is not statistically corrected. We show the specific heat for both the CRBM and FRBM at $L=8$ in  Fig.~\ref{fig:kitaev_MC_L=8_nocorr} and the LCRBM at $L=24$ in Fig.~\ref{fig:kitaev_MC_L=24_nocorr}. Observe that the CRBM and LCRBM without corrections are close to the corrected versions while the FRBM results lie quite far. This is due to the fact that the FRBM was no able to learn the interaction in the z-direction. Furthermore, note that the increase in lattice size did not affect the closeness between corrected and uncorrected CRBM.

%% file: Aanalytical.tex
\label{sec:Aanalytical}
As an alternative to learning the 2D Ising model, it is possible to compute the values of the convolutional kernels analytically. This is done similarly to Ref.~\cite{IsingAnalytical}.
The energy associated with an interaction between two neighboring spins in the Ising model is $E(s_1,s_2)=s_1s_2$. For the CRBM if the kernel is chosen to be:
\begin{align*}
W_1=&
\begin{pmatrix}
W & 0 \\
W & 0 \\
\end{pmatrix},\\
W_2=&
\begin{pmatrix}
W & W \\
0 & 0 \\
\end{pmatrix},\\
\end{align*}
the interaction term between $s_1$ and $s_2$ is $F(v_1, v_2)=\frac{v_\text{bias}(v_1+v_2)}{4}+\log(1+e^{h_\text{bias} + W(v_1+v_2)})$.
Note that for each $v_i$, the $v_\text{bias}v_i$ term will appear 4 times in the free energy. This produces the $\frac{1}{4}$ factor in front of $v_\text{bias}$. Also, note that $s=\pm 1$ and $v = \{0,1\}$. If $F(v_1, v_2)=\beta E(s_1, s_2)-C$,  then the CRBM will\textbf{} be equivalent to Ising model. $s_1,s_2$ can take four different states, which yield three equations. Eq.~\ref{eq:C1} is obtained for the case where both spin are up. Eq.~\ref{eq:C2} is obtained for the case where one spin is up and the other is down, and Eq.~\ref{eq:C3} for the case when both spins are down.
\begin{align}
\frac{v_\text{bias}}{2}+\log{1+e^{h_\text{bias} + 2W}}&=\beta+C, \label{eq:C1}\\
\frac{v_\text{bias}}{4}+\log{1+e^{h_\text{bias} + W}}&=-\beta+C,\label{eq:C2}\\
                                       \log{1+e^{h_\text{bias}}}&=\beta+C,\label{eq:C3}
\end{align}
they can be reduced to:
\begin{align}
v_\text{bias}&=2\log{\frac{1+e^{h_\text{bias}}}{1+e^{h_\text{bias} + 2W}}}, \\
2\beta&=\log{\frac{\sqrt{(1+e^{h_\text{bias}})( 1+ e^{h_\text{bias} + 2W})}}{1+e^{h_\text{bias} + W}}},\label{eq:C5}
\end{align}
and Eq.~\ref{eq:C5} can be re-expressed as:
\begin{align}
h_\text{bias}=-2 W + \log \left(\frac{\left(e^W-1\right) \left(\pm\sqrt{\left(e^W+1\right)^2-4 e^{4 \beta+W}}\right)-2
   e^{4 \beta+W}+e^{2 W}+1}{2 \left(e^{4 \beta}-1\right)}\right).
\end{align}
It only has a real solution if the term in the square root is positive, which means that:
\begin{align}
    |W| > W_\text{min}=\log \left(2 e^{2 \beta} \left(\sqrt{e^{4 \beta}-1}+e^{2 \beta}\right)-1\right).
\end{align}
We expect that the minimal $W$ would yield the smallest autocorrelation time. For this case:
\begin{align}
    W             &= \pm W_\text{min}, \nonumber \\
    h_\text{bias} &= -W \nonumber, \\
    v_\text{bias} &= -2W.
\end{align}
Using this analytical result for performing MC yields similar autocorrelation times and the same expectation values as obtained from a trained kernel.

%% file: Atrainingandsampling.tex
\label{sec:A_parameters}

In this Appendix, we list all our training and sampling parameters.

\paragraph{Training}
All models were trained with ADAM with a learning rate of $\lambda=10^{-3}$ and a batch-size of 20.
The Ising model was trained at $L=3$. First, the energy $E_\text{phys}(x)$ of all possible $2^{3^2}=512$ states was computed. Then the loss in Eq.~\ref{eq:loss} was minimized until a loss of $10^{-7}$ is reached. This procedure is demonstrated in a Jupyter Notebook in \cite{colab}.
\\
The Kitaev model was trained at $L=8$ in two steps. First, the pre-training step described in App.~\ref{sec:Atraining} was performed until the loss did no longer decrease, this happened on average after 800 epochs. Second, $10^4$ states are sampled with Metropolis MC. These states are then combined with a buffer of states sampled from the CRBM. The buffer update size is set to $ub=200$ as described in App.~\ref{sec:Atraining}. This step is iterated until the loss no longer decreases, which happens on average after 30 epochs.
\\ \\
\paragraph{Sampling}
For the Ising model sampling for local Metropolis was done on the CPU and sampling for the CRBM was done on the GPU. A local Metropolis step takes $t_\text{metro}=55$ns on a Intel(R) Core(TM) i7-9700K CPU. The time it takes to perform a CRBM Gibbs update step is lattice size-dependent, the ratio between a Metropolis and a CRBM update step $k=\frac{t_\text{CRBM}}{t_\text{metro}}$ is computed for a GeForce RTX 2070 Super and is shown in Fig.~\ref{fig:timeCRBM}. To compare the speed of both methods $k$ Metropolis steps are performed between each recorded state, whereas for the CRBM only one Gibbs update step is performed. The different temperatures are simulated after each other starting with the highest temperature. For both CRBM and Metropolis MC a warm-up phase is performed before the recording of the observables. For consistency, this warm-up phase was chosen to be $10\%$ of the total simulation.
\\
To perform the local Metropolis sampling for the Kitaev model, parallel tempering was used. A Metropolis update at each temperature and a temperature exchange are performed iteratively.
To sample using the CRBM three steps are performed repeatedly. First, at each temperature states are updated $k$ times using the Gibbs update from the CRBM. Second, a correction step where the CRBM and physical distribution exchange states is performed. Note that for this, the free energy of the Kitaev model needs to be computed. Lastly, a normal temperature exchange is performed. The number of steps $k$ is varied depending on $L$. $k$ is chosen such that the Gibbs sampling of the CRBM takes as much time as one correction step. Note that the most expensive operation in the correction step is the computation of the free energy. The timings and values for $k=\frac{t_\text{Kitaev}}{t_\text{CRBM}}$ can be found in Tab.~\ref{tab:timings}. Note that the warm-up for Metropolis MC was chosen to be $50\%$ of the generated samples, and the one for the CRBM just $10\%$. The higher percentage for Metropolis was chosen due to the higher autocorrelation time. 
\begin{minipage}{.475\textwidth}
\begin{table}[H]
    \begin{tabular}{|l|c|c|c|c|c|}
      \hline
      $L$                & 8      & 16     & 20     & 24     & 30\\ \hline
      CRBM Gibbs         & 0.09ms & 0.14ms & 0.17ms & 0.21ms & 0.28ms\\
      Kitaev free energy & 0.35ms & 4.5ms  & 12ms   & 30ms   & 98ms\\ \hline
      $k$                & 4      & 30     & 70     & 140    & 350 \\ \hline
    \end{tabular}
    \caption{Speed of one CRBM Gibbs update with a $2\times5\times5$ kernel and the computation of the free energy of the Kitaev model with an Intel(R)~Xeon(R)~X5670 at different lattice sizes. $k$ is the amount of Gibbs updates that are performed between correction steps.}
    \label{tab:timings}
\end{table}
\end{minipage}%
\begin{minipage}{.05\textwidth}
 $\quad$
\end{minipage}%
\begin{minipage}{0.475\textwidth}
\begin{figure}[H]
	\centering
	\includegraphics[width=0.7\textwidth]{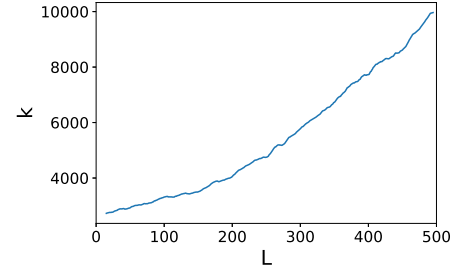}
	\caption{Ratio $k=\frac{t_\text{CRBM}}{t_\text{metro.}}$ between the time it takes to perform one local Metropolis step in the Ising model and the time to perform one CRBM Gibbs update step with a kernel of size $2\times2\times2$.}
	\label{fig:timeCRBM}
\end{figure}
\end{minipage}